\documentclass{aastex631}

\usepackage{graphicx}

\usepackage{dcolumn}
\usepackage{bm}

\usepackage{rotating}
\usepackage{subfigure}
\usepackage{amsmath}
\usepackage{hyperref}
\hypersetup{
    colorlinks=true,
    linkcolor=blue,
    filecolor=magenta,      
    urlcolor=red,
    pdftitle={Overleaf Example},
    pdfpagemode=FullScreen,
    }
\usepackage{color}
\hypersetup{
    colorlinks=true,
    linkcolor=red,
    citecolor=blue,
}

\begin{document}


\title{Identification and modelling of optically thin inverse Compton scattering in the prompt emission of GRB131014A}

\author{Pragyan Pratim Bordoloi}
\email{pragyan.bordoloi21@iisertvm.ac.in}
\affiliation{%
 School of Physics, Indian Institute of Science Education and Research Thiruvananthapuram, Kerala, 695551, India\\
}
\author{Shabnam Iyyani}%
\email{shabnam@iisertvm.ac.in}
\affiliation{%
 School of Physics, Indian Institute of Science Education and Research Thiruvananthapuram, Kerala, 695551, India\\
}%
 \affiliation{%
 Centre of High Performance Computing, Indian Institute of Science Education and Research Thiruvananthapuram, Kerala, 695551, India\\
}

\begin{abstract}

The mechanism responsible for the prompt gamma-ray emission of a gamma-ray burst continues to 
remain an enigma. The detailed analysis of the spectrum of GRB 131014A observed by the {\it Fermi} 
gamma ray burst monitor and Large Area Telescope 
has revealed an unconventional spectral shape that significantly deviates from the typical Band 
function. The spectrum exhibits three distinctive breaks and an extended power law at higher 
energies. Furthermore, the lower end of the spectrum aligns with power-law indices greater 
than -0.5, and in the brightest region of the burst, these values approach +1. The lowest spectral break is thereby found to be consistent with a blackbody. These observed 
spectral characteristics strongly suggest the radiation process to be inverse Compton 
scattering in an optically thin region. Applying the empirical fit parameters for physical 
modeling, we find that the kinetic energy of the GRB jet of bulk Lorentz factor, $\Gamma \sim 400$, 
gets dissipated just above the photosphere, approximately at a radius of $\sim 10^{14}$ 
cm. The electrons involved in this process are accelerated to a power-law index of $\delta = -1.5$, and the minimum electron Lorentz factor, $\gamma_{min}$, is approximately $3$. In summary, 
this study provides a comprehensive identification and detailed modeling of optically thin inverse Compton scattering in the prompt 
emission of GRB 131014A.

\end{abstract}

\section{Introduction}

Gamma ray bursts are the brightest explosive transients occurring in the distant cosmos. The origin of the intense 
gamma rays produced during the event is still debated. The uniqueness and non-repeating nature of GRB along with its wide variety of emission light curves makes it challenging to develop a generic radiation model for GRBs.  

Within the classical fireball model scenario \citep{Meszaros2006,Kumar_Zhang2015,Iyyani_2018}, the observed gamma ray emission is anticipated to be produced from the photosphere 
and in the optically thin regions above the photosphere. The high optical depth leading to the numerous scatterings of the radiation with the plasma results in thermalised emission from the photosphere. 
However, the kinetic energy dissipated in the site above the photosphere leads to 
relativistic shocks wherein the accelerated electrons cool via various non-thermal radiation processes. The competing non-thermal emission mechanisms are synchrotron emission \citep{Rees_Meszaros1992,Tavani1996,Sari_Piran1997,Beniamini_Giannios2017,Beniamini_etal_2018} and Inverse Compton scattering (ICS) \citep{Panaitescu_Meszaros2000,Stern2004,Peer&Waxman2004,Nakar_etal_2009,Ahlgren_etal_2015,Iyyani_etal_2015}. 

Synchrotron emission has long been favored as a model to explain non-thermal emissions in astrophysical phenomena. 
While it is widely accepted for producing non-thermal spectral shapes, straightforward synchrotron emission models struggle to account for the prompt emission spectra of GRBs. 
These challenges include hard low-energy spectral slopes, a narrow range of spectral peak energies \citep{Iyyani_2018}, and the majority of GRB spectra being consistent with slow-cooled synchrotron 
emission models have lower radiation efficiencies, while, the fast-cooled synchrotron emission, is radiation-efficient, however, produces broad spectral peaks that are 
found to be inconsistent with the data \citep{Burgess2014a} etc. Nonetheless, several modified versions of synchrotron emission scenarios have been proposed to address these limitations \citep{Dermer_etal_2000,Asano_etal_2009,Uhm_Zhang2014,Beniamini_etal_2018,burgess_etal_2019NatAs}. 

The alternative non-thermal emission process that has been actively studied is inverse Compton scattering under 
various scenarios such as subphotospheric dissipation \citep{Ghisellini&Celotti1999,Iyyani_etal_2015,Ahlgren_etal_2015} and synchrotron self Compton emission models in 
context of prompt emission \citep{Granot_etal_2000,Lloyd_Petrosian2000,Stern2004,Zhang_Yue_etal_2019} and afterglow emissions \citep{ssc1_2019,ssc2_2019,ssc3_2019,ssc4_2001,ssc5_2020}. 
A significant limitation in this model, as highlighted by \citealt{Piran_etal_2009}, arises when the seed photons are considered at lower energies, such as optical or infrared. If the observed gamma-ray emission is considered 
to result from the inverse Compton upscattering of these seed photons, it implies a substantial amplification or, in 
other words, a high average electron Lorentz factor. This, in turn, would lead to higher order scatterings with 
emissions peaking at TeV energies, ultimately straining the energy budget of a GRB. However, this issue can be 
addressed if the seed emission originates at relatively higher energies such that the amplification reduces. Moreover, if the upscattered photons surpass the 
energy threshold for pair production, this can subsequently reduce the average electron Lorentz factor, thereby suppressing the higher-order scatterings \citep{Piran_etal_2009}.

In this study, we characterize an uncommon spectral shape with multiple spectral breaks in GRB 131014A through a comprehensive spectral analysis. Additionally, we model the observed spectrum using the optically thin inverse Compton 
scattering radiation mechanism. In sections \ref{obs} and \ref{spectral_analysis}, we describe the observations and the spectral analysis respectively. In section \ref{physical_modeling}, we 
present the physical modelling to derive the physical scenario of the observed spectra. We present the 
discussions on the observed spectral properties and its physical interpretation in section \ref{discussion} and finally conclude in section \ref{conclusion}.

\section{Observations}  \label{obs}

On 14 October 2013, multiple space satellites including {\it Fermi} Gamma-ray space telescope 
\citep{Fermi_GBM_131014A,Fermi_LAT_131014A}, Konus Wind \citep{Konuswind_131014A} and Suzaku Wide-band All-sky 
Monitor (WAM; \citep{Suzaku_131014A}) triggered on the GRB131014A. The burst was localised in the sky at RA = $100.29^{\circ}$ and Dec = 
$-19.13^{\circ}$ with nearly $53'$ uncertainty \citep{IPN_131014A}. Nearly 12 hours post the trigger, the 
target of opportunity (ToO) observations carried out by Swift X-Ray Telescope (XRT) resulted in a detection of a fading X-
ray source \citep{Swift_XRT_13104A}. An optical transient tentatively linked to this event was also later reported \citep{NOT_OT_131014A,GROND_OT_131014A,Swift_OT_upperlimit_131014A}. However, there was no robust redshift measurement made for the burst. 
\begin{figure}[!ht]
    \centering
    \includegraphics[width=1.0\textwidth]{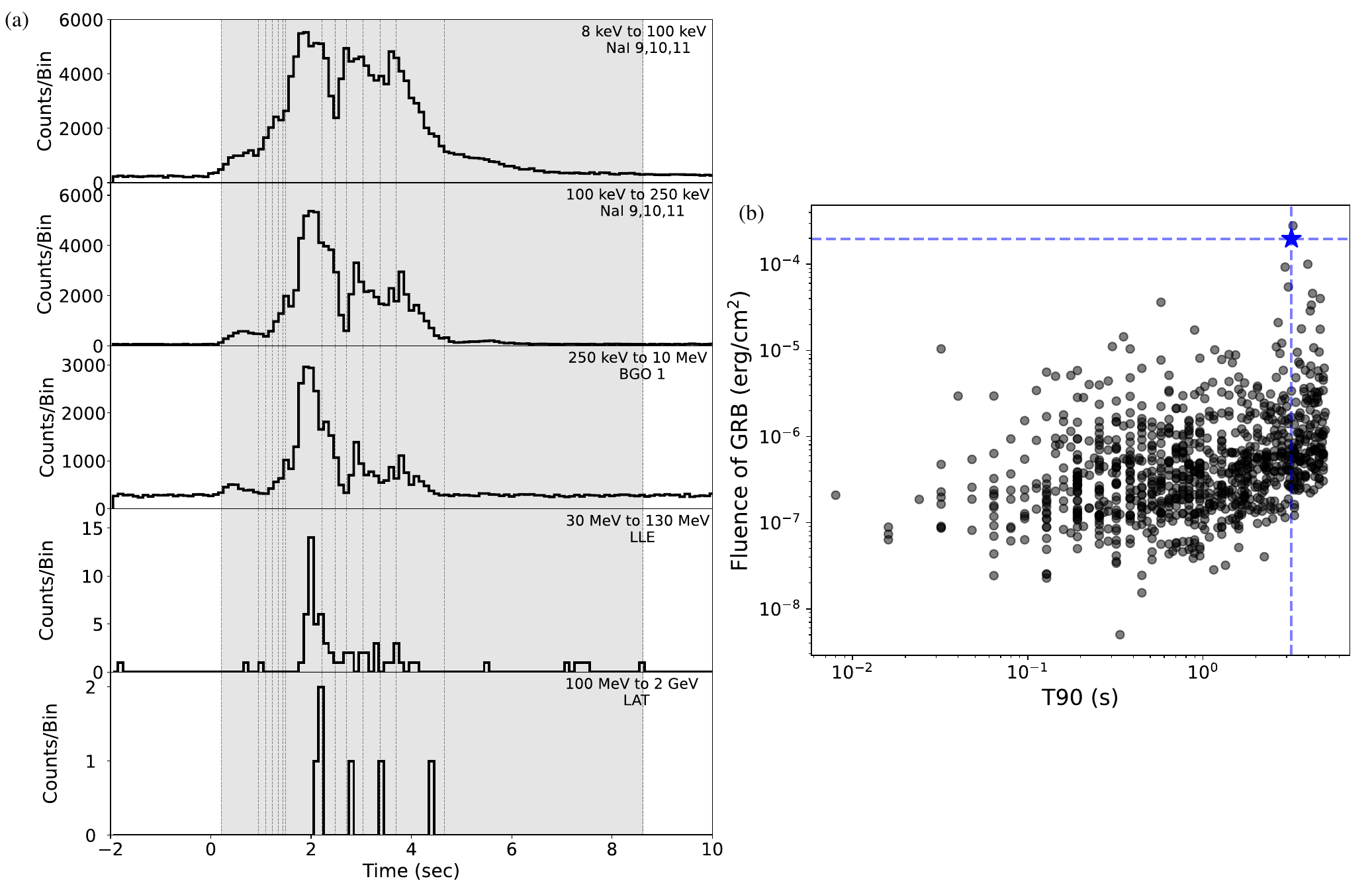}
    \caption{(a) The multi-panel light curve (black solid line) of GRB 131014A with $0.1$ s binning, as observed across different energy ranges of NaI(9,10,11), BGO 1 and LAT detectors is shown. 
    The shaded grey area represents the duration of the burst studied for time resolved spectral analysis. Additionally, the dotted vertical lines depict the time resolved intervals obtained from Bayesian block binning. (b) The scatter plot of fluence versus $T_{90}$ for {\it Fermi} GBM-detected GRBs with $T_{90} \le5 \, \rm s$ is shown, with the blue star indicating GRB 131014A.}
    \label{lightcurve}
\end{figure}
At 05:09:00.20 UT ($T_0$), the GRB triggered the two instruments aboard the {\it Fermi}  spacecraft: the Gamma Ray Burst Monitor (GBM, covering 8 keV 
to 40 MeV) and the Large Area Telescope (LAT, spanning 30 MeV to 300 GeV). 
The burst displayed a complex, multi-pulsed light curve, depicted in Figure \ref{lightcurve}a, illustrating its 
activity across increasing energy levels. The lightcurve includes observations of various detectors aboard GBM, including sodium iodide (NaI) covering 8 keV - 100 keV and 100 keV - 250 keV, as well as bismuth germanate (BGO) covering 250 keV - 30 MeV. Additionally, data from the LAT Low Energy (LLE, 30 MeV - 130 
MeV) and LAT ($> 100 , \rm MeV$) detectors are included. The LLE and LAT emission commenced at $T_0 + 0.06 \, s$ and $T_0 + 2.16 \, s$ respectively. The LAT observations extended until $T_0 + 15 \, s$ and detected a 
total of $31$ photons with energies exceeding $100 \, \rm MeV$, with the highest photon energy recorded at $1.8 \, \rm 
GeV$ occurring at $T_0 + 14 \, s$ \citep{Fermi_LAT_131014A}, while, the highest LAT photon energy observed during the GBM emission phase is $1.23$ GeV at $T_0 + 2.183$ s.
The burst had a duration of $T_{90} = 3.2 \pm 0.09 \, \rm s$ in 50–300 keV range where $T_{90}$ represents the time interval during which 
$90\%$ of the burst fluence is measured. The burst recorded an energy fluence of $1.98 \times 10^{-4}\, \rm erg\, 
cm^{-2}$ in 10 - 1000 keV, positioning it as the second most luminous among the shorter duration GRBs\footnote{Based on the Fermi GBM GRB spectral catalog available at https://heasarc.gsfc.nasa.gov/W3Browse/fermi/fermigbrst.html as of May 2025.} observed by {\it Fermi} with 
$T_{90} \le 5\, s$ (Figure \ref{lightcurve}b). 

\section{Spectral Analysis}  \label{spectral_analysis}
The spectral analyses were conducted using the Multi-Mission Maximum Likelihood (3ML, \citealt{3ml}) package. The analysis included {\it Fermi} data from the three brightest NaI detectors, NaI 9, 10 and 11 with source angles $< 60^{\circ}$ \citep{GBMcatalog2014} along with BGO 1 and LAT detectors 
providing a spectral coverage from $8$ keV to several GeV. In case of NaI detectors, data in the energy range 30 keV to 40 keV corresponding to the iodine-K edge were excluded, in addition, to those in the extreme edges such as those below 8 keV and those above 900 keV. For BGO, LAT-LLE 
and LAT ($> 100 \, \rm MeV$) detectors, the data within the energy ranges 250 keV - 10 MeV, 30 MeV - 100 MeV and 100 MeV - 2 GeV were used 
respectively. The burst interval from 0 to 8.6 s where chosen for both the time integrated and time resolved analyses. We note that LAT data ($> 100\, \rm MeV$) is not available for analysis in all the time-resolved intervals. However, in the intervals where LAT data is available, the spectrum is analysed jointly over the energy range of 8 keV to 2 GeV.
The background was modelled using a polynomial function best fitted to data in the intervals pre ($T_0 - 20$s to $T_0 - 2$s) and post ($T_0 + 20$s to $T_0 +30$s) 
the burst duration. The spectra are fitted using a Poisson–Gaussian likelihood, which accounts for the Poisson-distributed total counts and the Gaussian-distributed background estimates \citep{Burgess_etal_2019}. The corresponding negative logarithm of the likelihood ($-log({\cal{L}})$) is referred to as Pgstat\footnote{https://heasarc.gsfc.nasa.gov/xanadu/xspec/manual/XSappendixStatistics.html}. The technique of maximum likelihood estimate was employed for estimating the model parameters while the Akaike Information Criterion\footnote{AIC = $2k\, - \,2log({\cal{L}})$ where $k$ is the number of free parameters of the model.} (AIC, \citealt{Akaike1974}) was used for the model selection.       


\begin{figure}[!b]
  \centering
   \includegraphics[width=1.0\textwidth]{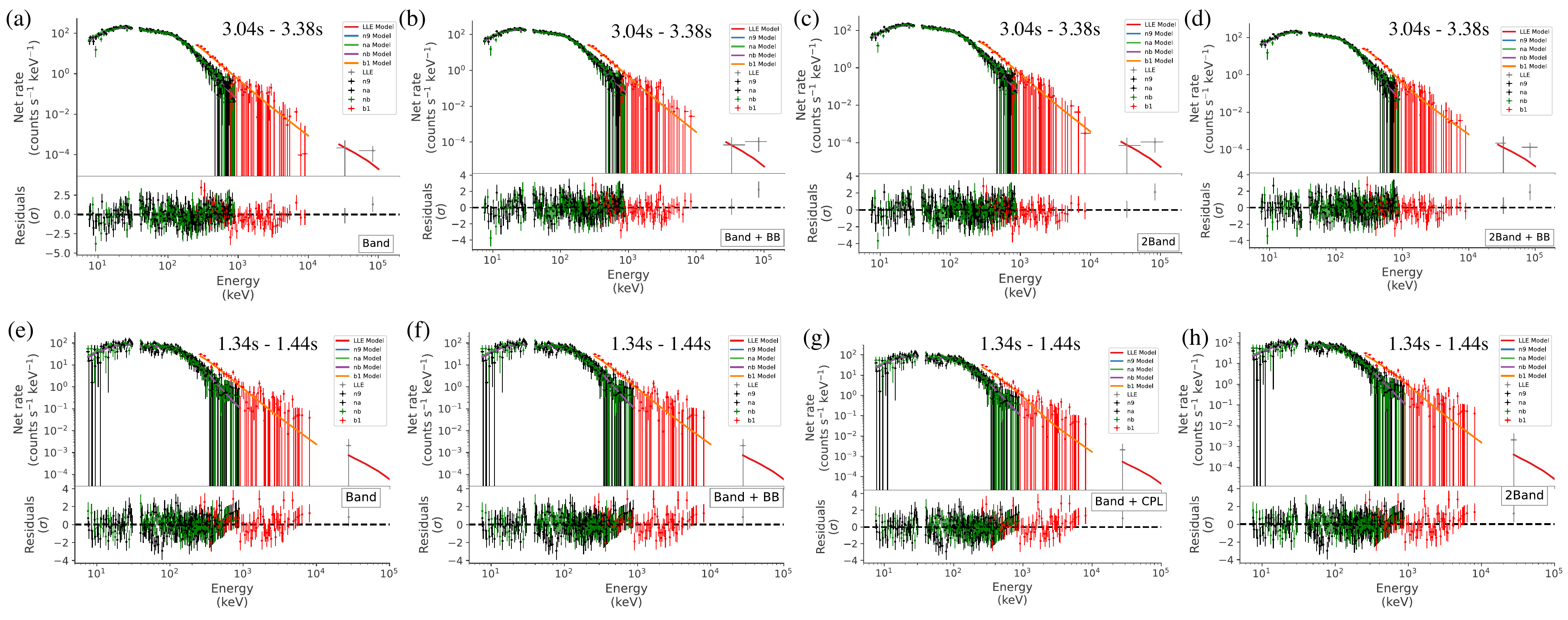}
        \caption{The counts spectra and residuals for different spectral models fitted to two time intervals are shown.   Panels (a–d) show fits for [3.04 s, 3.38 s] using Band, Band + BB, 2Bands, and 2Bands + BB.  Structured, wavy residuals are apparent in the first three models, indicating model inadequacy. In contrast, the 
        residuals for the 2Band + BB model show no such pattern, appearing more randomly scattered around 
        zero — consistent with a well-fit model. This visual evidence supports the statistically significant $\Delta$AIC improvement reported in Table \ref{table_fit}. Panels (e–h) show fits for [1.34 s, 1.44 s] using Band, Band + BB, Band + CPL and 2Bands models, respectively. Residuals appear largely random, consistent with comparable $\Delta$AIC values (Table \ref{table_fit}), indicating all models are equally plausible. Note LAT data ($> 100\, \rm MeV$) is  not available in these time intervals.} 
    \label{residuals}
\end{figure}

The spectral analysis allows to identify and characterise the shape of the spectrum and subsequently the emission mechanism. Given the dynamic nature of GRB emissions, evident from their intricate light curves, 
employing the time-resolved spectroscopy becomes crucial. This approach enables us to identify the underlying instantaneous spectral shape, which might otherwise get smeared, leading to lose of precise spectral information when modeled over the entire duration during time-integrated analysis. Furthermore, in 
order to investigate the temporal evolution of the spectrum, a time-resolved spectral analysis of the burst was conducted. The time intervals of the analysis were determined using the Bayesian block binning method \citep{scargle_1998} which resulted in 14 time bins. A detailed time resolved spectroscopy was carried out for the burst duration from 0 s to 8.6 s.

We further note that the primary goal of the detailed time-resolved spectral analysis is to determine the statistically preferred spectral shape while ensuring physical 
consistency across the burst emission. This approach facilitates a comparative study of the evolution of various spectral parameters, aiding in the development of a physical model for the 
underlying radiation process. Thus, the best fit model reported for this GRB across various time intervals are after taking into account both lowest AIC and the model that offers greater physical consistency.   

The underlying instantaneous spectral shape of the burst radiation can be best identified in the time resolved brightest bins with higher number of counts. Thus, the spectral analysis was initiated in the time intervals during the bright regions of the burst assuring enough number of photon counts for analysis. 
The joint spectral analysis including data from GBM and LAT detectors requires an effective area correction to be incorporated. 
The spectral data was initially modelled using the conventional models such as the Band function \citep{Band_1993} 
alone and in combination with a blackbody (Band + BB) \citep{Axelsson2012,Iyyani2013}. In addition, several 
intermediate complex models such as `Smoothly Broken Power Law (SBPL)', `Blackbody + Band $\times$ High 
Energy Cutoff', `Cutoff Power Law + Power Law', `Cutoff Power Law (CPL) + Blackbody', `Band + Cutoff 
Power Law', `Band + Cutoff Power Law + Power Law', `Band + Cutoff Power Law + Blackbody', `2SBPL' (in reference to studies such as \citealt{Ravasio_etal_2018,Ravasio_etal_2019A,Toffano_etal_2021}) and a few more, were also tried for spectral modelling.
However, these models 
resulted in poor fits with wavy residuals, often with unconstrained parameters, and occasionally with spectral components switching places between consecutive fits leading to inconsistent temporal evolution of the parameters. This strongly suggested that the spectrum has a more complicated shape. 

A significant improvement in fitting was observed by adding an extra Band function to the Band + BB model, resulting in the most random 
structure of the residuals. We note that while the Band + CPL + BB model is an equally good fit in some 
time intervals, it results in poor fits in others. Despite having one fewer free parameter compared to the 2Bands + BB model, the Band model is more flexible than the Cutoff Power Law, allowing it to more accurately capture the 
spectral shapes of other time-resolved intervals as the spectrum evolves. Thus, the 2Bands + BB model is found to better capture the overall spectral evolution (see Table \ref{table_fit} in Appendix \ref{appendix1}).

By choosing 2 Bands + BB as the best fit model, the effective area correction factors i.e, the normalisation offsets of the detectors were estimated by fitting the data with this spectral model 
multiplied by a constant of normalisation for each instrument. The constant of the brightest NaI detector i.e NaI 10 was frozen at unity while that of all other detectors were kept free. The effective area correction factors for the various detectors were found as follows: $1 \pm 
0.009$, $1 \pm 0.009$, $1.03 \pm 0.015$ and $1.8 \pm 0.6$ for NaI 9, NaI 11, BGO 1 and LLE respectively. We note that the effective area correction factor estimate for LLE is relatively high. To address this, we fixed the correction value for LLE at $1.2$, which is within the error margin of the initial estimate for our analyses. Including the effective correction factors, the best fit model, 2 Bands + BB model brought about an improvement in the AIC statistics\footnote{The definition of $\Delta AIC$ and its interpretation is presented in Appendix A.} \citep{Akaike1974} by $\Delta \, \rm AIC = 49$,  $\Delta \, \rm AIC = 28$ and $\Delta \, \rm AIC = 16$  with respect to the Band function fit alone, Band + BB and Band + CPL + BB respectively in the time interval [3.04s, 3.38s]. 
To compare the spectral model fits, the counts plots along with the residuals for Band-only, Band + Blackbody, 2Bands and 2Band + Blackbody models during the time interval [3.04s, 
3.38s]—where the most significant improvement in $\Delta \rm AIC$ is observed—are shown in Figure 
\ref{residuals} (a-d). The counts plots along with residuals for other models, including Band + Cutoff Powerlaw, Band + Cutoff Powerlaw + Blackbody and 2SBPL, are presented 
separately in the Figure \ref{Residual_2} of the Appendix \ref{appendix1}. Notably, the residuals for 
the 2Bands + Blackbody model exhibit most random behavior with minimal waviness. Note the LAT data ($>$ 100 MeV) are not available for this particular time interval.

\begin{figure*}[!ht]
    \centering
    \subfigure[Time interval : 1.34 s - 1.44 s]{
        \includegraphics[width=0.40\textwidth]{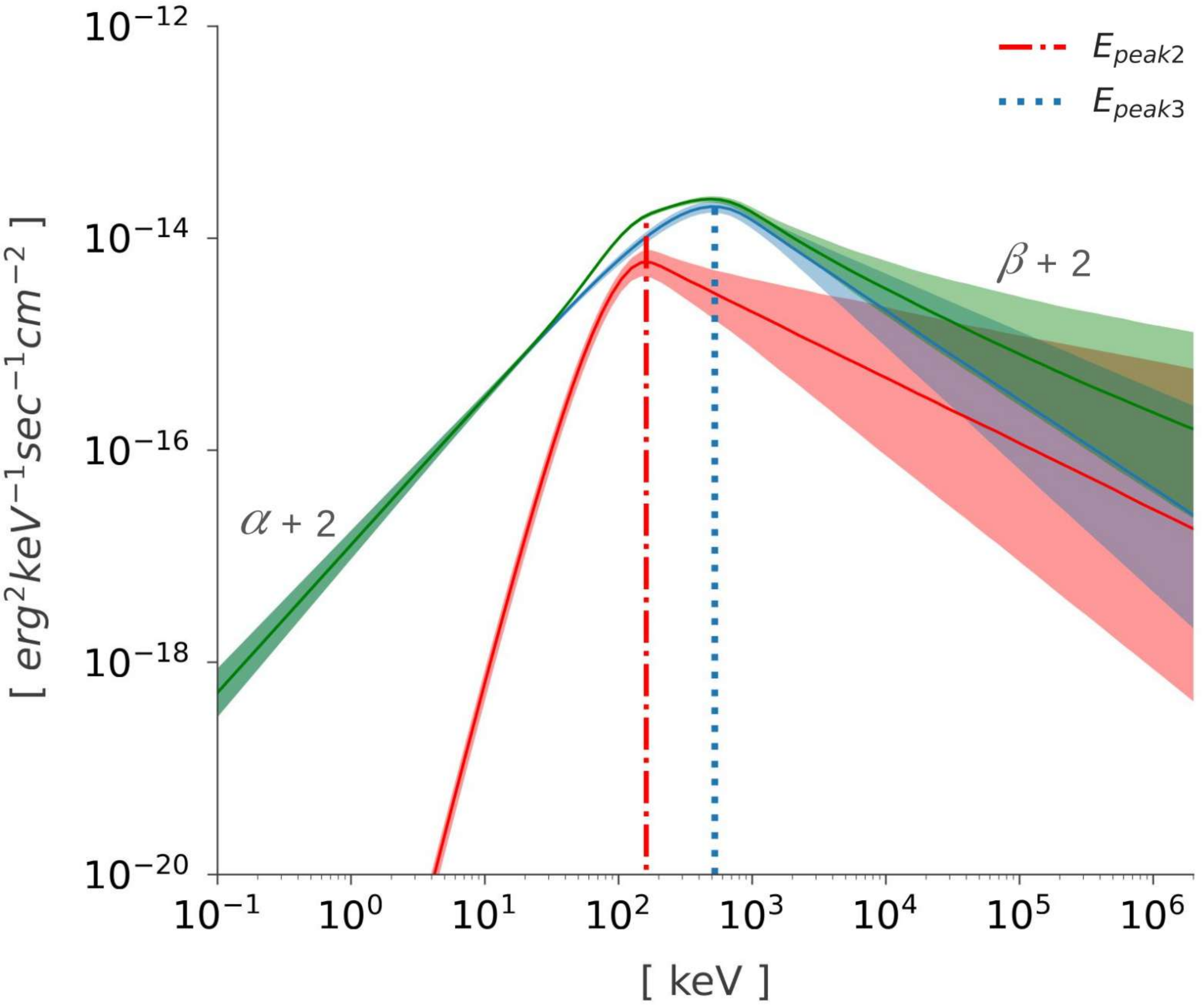}
    }
    \hspace{0.15in}
    \subfigure[Time interval : 3.04 s - 3.38 s]{
        \includegraphics[width=0.40\textwidth]{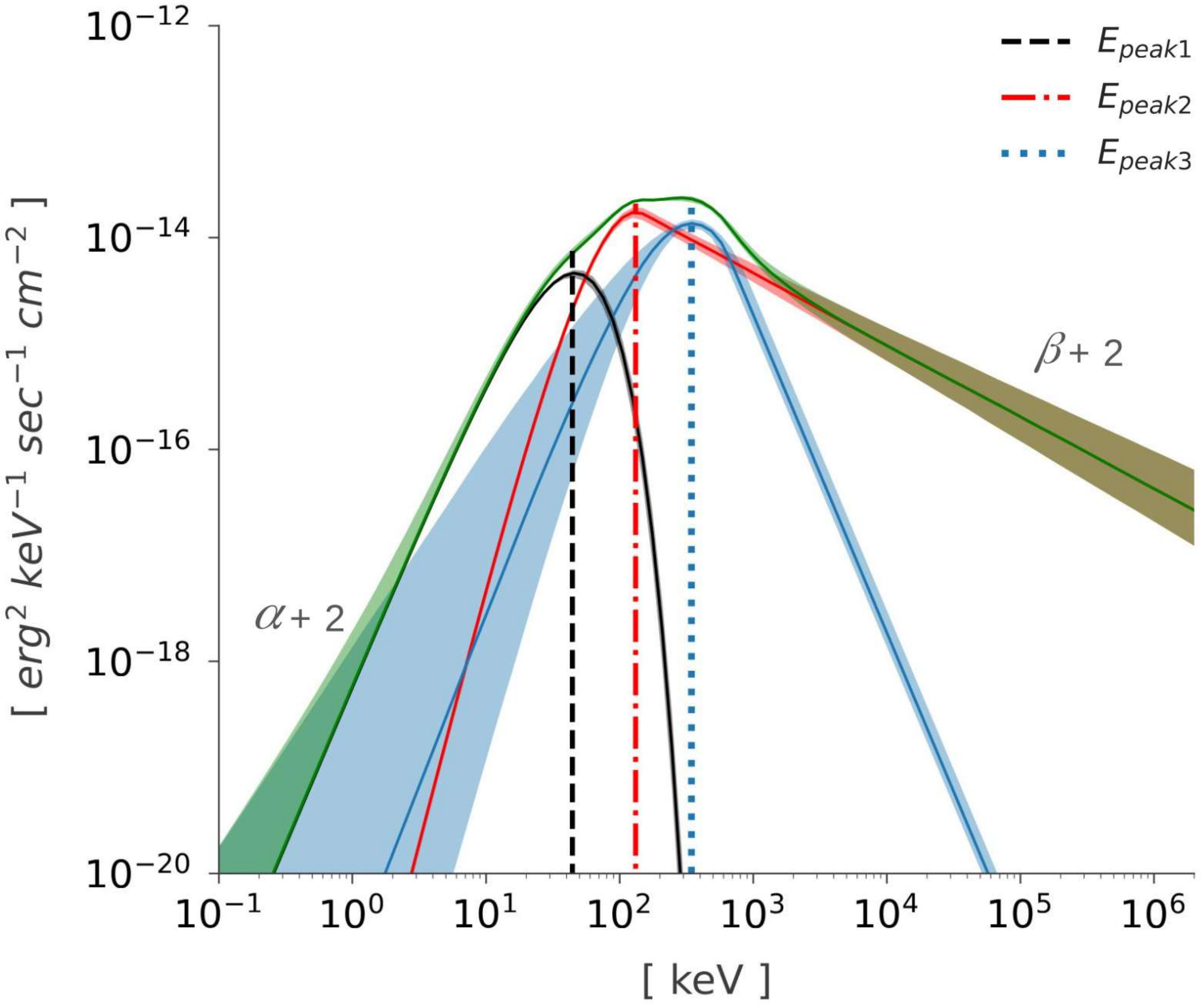}
    }
    \caption{The spectral shape of the burst spectrum characterised by the best fit empirical model: (a) 2 Bands in time intervals less than $1.49s$ and (b) 2 Bands+ Blackbody in the time intervals above $1.49s$ are shown in the above $\nu \, F_{\nu}$ plots in green solid line. The shaded green region signifies the $68\%$ confidence interval of the overall model. The resulting shape of the spectrum is defined by the parameters: $\alpha$ and $\beta$ represent the low and high energy asymptotic power law indices respectively, while the three $\nu \, F_{\nu}$ peaks denoted by the $E_{peak1}$, $E_{peak2}$ and $E_{peak3}$ are marked by black (dashed), red (dash dotted) and blue (dotted) lines respectively. The individual spectral components of the model are also shown in black (Blackbody), red (Band$_1$) and blue (Band$_2$) solid lines along with their $68\%$ confidence intervals with same shaded colour. Note that the above $\nu F_{\nu}$ plots are shown for 0.1 keV - 2 GeV to illustrate the model’s extrapolated asymptotic behaviour in lower energies while the spectral data cover only 8 keV - 2 GeV.}
    \label{nuFnu}
\end{figure*}

We further note that during the time-resolved analysis, in the initial time bins from 0 s to 1.49 s, the blackbody component could not be well-constrained, making 2Bands the best fit model. However, we note that other models like 'Band', 'Band + BB', 'Band + Cutoff Power Law' was sometimes an 
equally good fit (Figure \ref{residuals} (e-h); refer Table \ref{table_fit}) while the overall spectral shapes obtained remained unchanged. For continuity in modeling throughout the burst 
emission, we chose 2Bands as the best fit model in the initial bins where the blackbody component could not be constrained. 
This choice also ensured the continuity of the spectral model and consistent temporal evolution of the parameters.
The identified shape of the overall spectrum (green solid line) of 
GRB131014A as in the $\nu F_{\nu}$ space in the regions before 
and after $1.49s$ are shown in Figure \ref{nuFnu}(a) and \ref{nuFnu}(b) respectively. The overall spectral shape exhibits multiple peaks/ breaks in the spectrum. 

In summary, the light curve (Figure \ref{lightcurve}a) indicates that significant LLE and LAT emission emerges around $1.5$ s, marking the onset of the brightest phase of the 
burst—from approximately 1.5 to 4 s—in terms of both energy output and photon count rate. In the time bin 1.49–2.22 s, where both LLE and LAT ($>100,\mathrm{MeV}$) data are 
available, the 2Bands and 2Bands + BB models yield statistically comparable fits (Table \ref{table_fit}, also see Figure \ref{Residual_2} a-g). However, the 2Bands + BB model is selected as the best-fit, as justified in Appendix A, given its physical consistency with subsequent intervals.
From 2.22 s onward, the BB component becomes statistically significant, indicating its increasing detectability as the burst brightens. This trend supports the interpretation that the BB component is likely present throughout the burst, but remains 
undetectable in the early phase due to lower temperatures and reduced flux levels (see Figure \ref{spec_param}c). As the photon statistics improve, so does the ability to 
resolve this component, as reflected in the systematically improving $\Delta$AIC values in Table \ref{table_fit}.

In the earlier phase (0–1.49 s), where LLE/LAT emission is minimal and the BB component remains unconstrained, the simpler 2Bands model yields the lowest AIC (not rejected by data) and is adopted for continuity. This choice ensures a smooth and physically meaningful evolution of spectral parameters, particularly the progression of the three 
$E_{\text{peak}}$ values, which closely track the burst intensity. Attempts to apply alternative models in these early bins resulted in inconsistencies and incoherent spectral peak evolution. The continuity provided by the 2Bands model in the early phase, and the 
clear preference for 2Band + BB during the brightest intervals, together support the interpretation that the complex spectral shape—including the three spectral peaks, and the extended power-law features—is not solely a consequence of LLE or LAT onset. Instead, 
these features are inherently present in the emission, but as the burst intensity increases i.e the spectral peaks shift to higher energies in tandem with the intensity, and the increased photon count 
enhances the statistical significance of underlying spectral features.

Furthermore, the time-integrated spectrum spanning from 0 to 8.6 s was also found to be well modelled using 2 Bands + BB yielding a significant improvement in the statistic by reducing the AIC by $229$, $79$ and $30$ with respect to the Band function only, Band + BB and Band + CPL + BB fits respectively. The obtained fit parameters of the best fit model is reported in the Table \ref{table_fit} in Appendix \ref{appendix1}. bf Thus, the spectral analysis of the burst reveals a unique overall spectral shape, distinct from the conventional GRB spectral shapes described by empirical functions like the Band function or Band + Blackbody. 

While we also note that one can always devise new empirical spectral 
models either a single mathematical function or by combining existing ones through additive or multiplicative forms, the critical aspect lies in the overall spectral shape and the physical 
interpretation presented to it. The robust affirmation of the interpretation would be a direct physical model fitting to the data (for example \citealt{Burgess2014a, Ahlgren_etal_2015,Ahlgren_etal_2019,Vianello2018}). In this work, 
the 2 Band + BB model is used solely as an empirical construct to capture the observed overall spectral shape, while our physical interpretation is guided by the key spectral features it reveals (see section \ref{section3.1}).

The burst fluence is estimated to be 5$\times$$10^{-4}$ $erg/cm^2$ which corresponds to a total burst energy of $E_{iso} = 2.7 
\times 10^{55}$ $erg$ assuming a redshift $z = 2.12$ \footnote{The average of the known redshifts of GRBs is reported to be around $z = 2.12$ \citep{racusin_2011}. It is also noteworthy that this redshift implies an unusually high $E_{iso}$, suggesting a redshift $z<2$ 
more plausible based on the typical distribution of isotropic-equivalent energies observed in GRBs \citep{Cenko2011}.}, radiation efficiency (1/Y) of 0.65 \citep{racusin_2011} and the standard $\Lambda_{CDM}$ cosmology along with the cosmological parameters, $H_0 = 67.4 \pm 0.5 \, km \, s^{-1} \, Mpc^{-1}$, $\Omega_m = 0.315$ and $\Omega_{vac} = 0.685$ \citep{Planck_2020}.

\subsection{Overall Spectral Shape Characterisation}
\label{section3.1}
 The multiple additive components composing the best fit spectral model 2 Bands + BB (Figure \ref{nuFnu}) are used here purely as an empirical tool to capture the complex overall spectrum, including its multiple breaks and the asymptotic power law behaviours observed at both lower and higher energies. This choice is not intended to advocate 2 Bands + BB as a new physical model, nor do we assign physical meaning to the individual Band components. Instead, the focus of this work is on the overall spectral shape resulting from the combined fit, which is what drives our subsequent physical interpretation provided in section \ref{section4}. 

It is noted that all the parameters of the individual components are not essential to characterise this net spectral shape\footnote{The parameter values obtained for the individual spectral components are provided in Table \ref{table_fit} in the Appendix to ensure reproducibility of the overall spectral shapes.}. Therefore, we parameterise the obtained overall spectral shape in terms of the following parameters:

\begin{itemize}
    \item $E_{peak1}$: The first spectral peak which is modelled by the blackbody component is referred to as $E_{peak1}$ which is around a few tens of keV. $E_{peak1}$ corresponds to the $\nu F_{\nu}$ peak of the blackbody fit. 
    \item $E_{peak2}$: The second spectral peak modelled by one of the Band functions is referred to as $E_{peak2}$ which is around a hundred keV.
    \item $E_{peak3}$: The third spectral peak modelled by the second Band function is referred to as $E_{peak3}$ with values around a few hundreds of keV. 
    \item $\alpha$: The low energy spectral index of the asymptotic power-law obtained below the spectral peak, $E_{peak1}$. 
    \item $\beta$: The high energy spectral index of the asymptotic power-law obtained above the spectral peak, $E_{peak3}$. 
\end{itemize}
The $\alpha$ and $\beta$ are determined by fitting a power law function to the low energy asymptotic part (below $E_{peak1}$) and the high energy asymptotic part (above $E_{peak3}$) of the photon flux plot of the overall spectral shape, 
respectively within the observation energy window (8 keV - 2 GeV). The $68\%$ confidence interval errors for these indices were derived 
by fitting the power law function to the low and high energy asymptotic parts of the upper and lower bound curves of the $68\%$ 
region of the overall photon flux spectral shape, respectively. These parameters are depicted on the $\nu F_{\nu}$ plot in the Figure \ref{nuFnu}.

Furthermore, it is interesting to note that in addition to the spectral peaks mentioned above, the asymptotic power-law in the higher energies creates an extra break. In other words, the asymptotic behaviour of the higher energy spectrum significantly deviates from the turnover predicted by the $E_{peak3}$ in the higher end. In the following subsection, the temporal evolution of these five parameters along with the energy fluxes are presented. 

\subsection{Spectral analysis results} 
\begin{figure*}[!ht]
    \centering
    \includegraphics[width=1.05\linewidth]{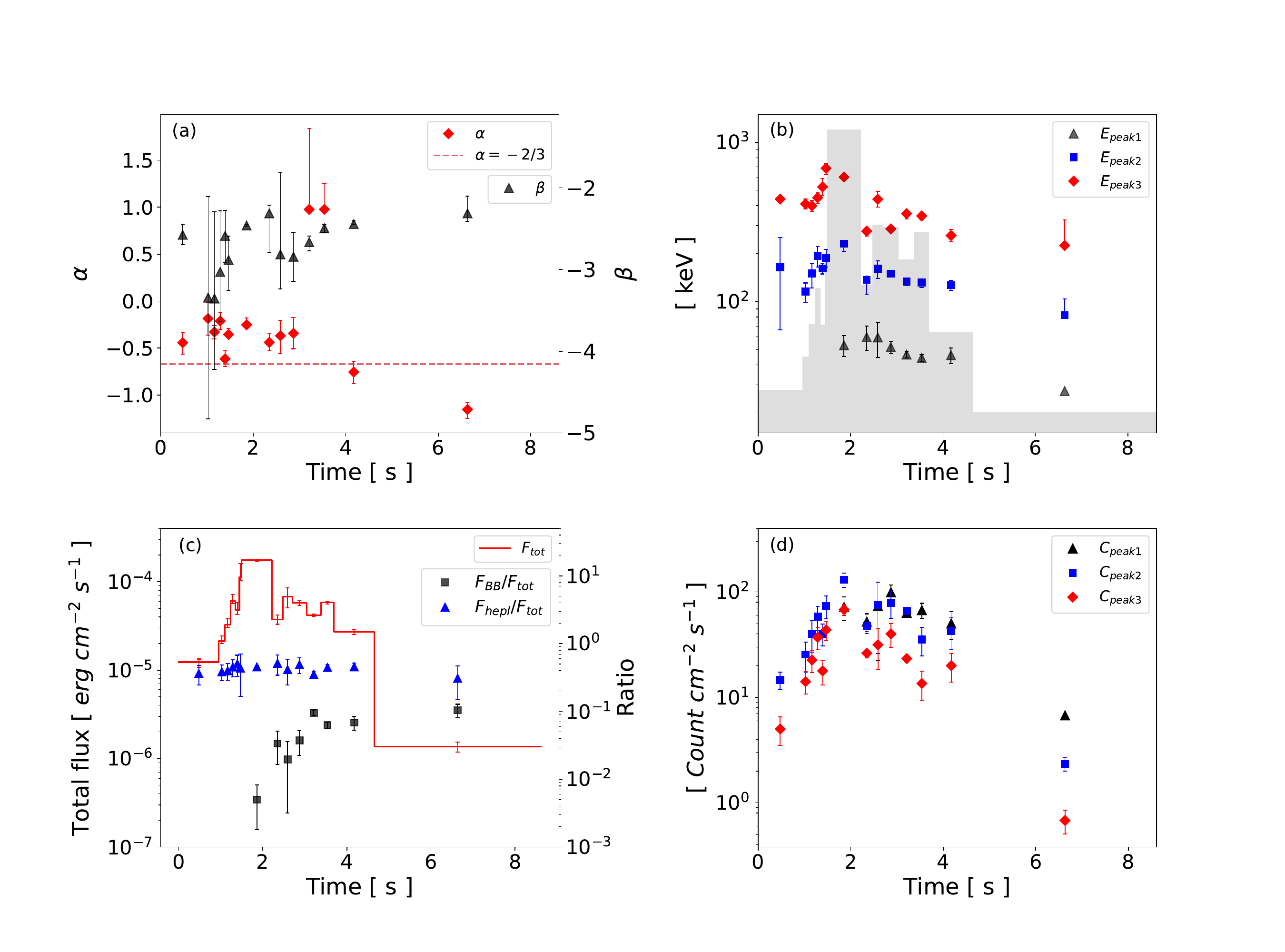}        
    \caption{ The temporal evolution of the model parameters characterising the spectral shape: (a) $\alpha$ (red diamond) and $\beta$ (black triangle); (b) spectral peaks, $E_{peak1}$ (black triangle), $E_{peak2}$ (blue square) and $E_{peak3}$ (red diamond) are shown. In plot (a), the dashed red line represents the 'line of death' ($\alpha = -2/3$) of synchrotron emission. In plot (b), the light curve is shown in shaded grey in the background for reference. (c) The temporal evolution of the total energy flux of the burst, $F_{tot}$, along with the ratios of the thermal energy flux ($F_{BB}$) and energy flux in the extended power law region of the spectrum, $F_{hepl}$ to the total energy flux are shown in red step line, black square and blue triangle respectively. Furthermore, in (d) the temporal evolution of the photon counts fluxes at the spectral peaks, $C_{peak1}$, $C_{peak2}$ and $C_{peak3}$ are shown in black triangle, blue square and red diamond respectively. 
    Note that since the BB component was not constrained until $1.49$ s, the corresponding parameters are absent in the aforementioned plots.}
    \label{spec_param}
\end{figure*}

The evolution of the spectral parameters of the 2Bands+BB model as characterised in the preceding section, obtained via the time resolved analysis is presented in the Figure \ref{spec_param}.
The time evolution of the parameter $\alpha$ shows no clear trend up to $\sim$3 s; instead, $\alpha$ exhibits a relatively flat evolution, with values predominantly harder than $-0.5$, after which it gets harder 
reaching values close to $+1$ (Figure \ref{spec_param}a). 
We note that $\alpha \sim +1$ is obtained in two consecutive time intervals (around $\sim 3\, \rm s$), which is atypical compared to the preceding 
evolution trend. Nevertheless, the overall behaviour remains consistent with a thermalisation trend at lower energies (see 
section \ref{section4.1} for further discussion). To verify the robustness of the fits in these bins, we performed the following checks: (i) replacing the BB component with a free-index CPL in the 2Bands + BB model yields $\alpha \sim +1.2$ for the [3.04–3.38] s 
and [3.38–3.69] s intervals, supporting the physical origin of the thermal component; (ii) the 2Bands + BB model remains statistically 
preferred ($\Delta$AIC; Table 1); and (iii) all three $E_{\mathrm{peak}}$ values exceed 20 keV, ensuring reliable constraints and ruling out low-energy 
artifacts \citep{Preece_etal_2016}. While these fits seem to be statistically sound, such a feature may arise from a specific 
physical scenario, minimal spectral smearing in these intervals (see section \ref{section4.1}), or potential limitations of the empirical spectral model fitted in this context.
As the burst progresses and the emission is integrated over longer time intervals particularly towards the end, $\alpha$ tends to become softer approaching values $\le -1$. Such broader integration windows can introduce spectral smearing, resulting in artificially softened $\alpha$ values compared to the true instantaneous spectral shape. 

The parameter $\beta$, which signifies the asymptotic high-energy power-law index of the spectrum, tends to exhibit a notably a steady trend, with values generally being around $-2.5$, as shown in Figure \ref{spec_param}a. 
Examining the three spectral peaks denoted by $E_{peak1}$, $E_{peak2}$, and $E_{peak3}$ over time, their evolution is found to track the burst's intensity, as illustrated in Figure \ref{spec_param}b where the light curve is plotted in the background for reference. 
Initially, the $E_{peak2}$ and $E_{peak3}$ increase with time, reaching their peaks at $1.47$ s and $1.85$ s, respectively, followed by a 
decreasing trend for the remaining duration of the burst. As the BB component is not constrained, $E_{peak1}$ is not characterised until $1.49$ s (see section \ref{non_detection_BB}), however, afterwards, the 
$E_{peak1}$ also shows a decreasing trend consistent with the other spectral peaks. We note that largely during the decaying phase, all the three spectral peak values tentatively follow a trend consistent with $t^{-0.6}$.
The temporal evolution of the total flux, $F_{tot}$ of the burst estimated for the observed energy range of the burst 8 keV to 2 GeV, is depicted in Figure \ref{spec_param}c. The $F_{tot}$ 
ranges between $10^{-5}$ to $10^{-4}$ erg/cm$^2$/s placing it among the brighter GRBs observed by {\it Fermi}. The flux light curve also demonstrates the multi-pulsed nature of the burst with several 
peaks. Furthermore, the temporal evolution of the ratios of the blackbody flux ($F_{BB}$) estimated in the energy range 8 keV - 5 GeV, and the energy flux within the extended power law at higher energies  ($F_{hepl}$) of the spectrum estimated in the energy range scaling from the $E_{peak3}$ till $5$ GeV, with respect to the $F_{tot}$ are also shown in Figure \ref{spec_param}c. 
The $F_{BB}/F_{tot}$ is found to increase with time from about $0.5\%$ to nearly $10.4\%$. On the other hand, the ratio $F_{hepl}/F_{tot}$ remains almost
constant at around $0.5$ throughout the burst duration, as shown in Figure \ref{spec_param}(c). 
In addition, the observed photon count fluxes obtained at the spectral peaks $E_{peak1}$, $E_{peak2}$ and $E_{peak3}$ are plotted in  black triangle, blue square and red diamond respectively in Figure \ref{spec_param}d. 
Throughout the burst, the average of the counts at the peaks $E_{peak2}$ and $E_{peak3}$ are generally less than those at $E_{peak1}$. 
Overall the counts at the spectral breaks peak at around $2.87$ s and subsequently decreases with time, closely following the total flux of the burst. 


\section{Physical Modelling} \label{physical_modeling}
\label{section4}
\subsection{Inconsistencies with synchrotron and sub-photospheric dissipation models}
\label{section4.1}
The non-thermal nature of the GRB spectrum is commonly attributed to different radiation mechanisms, such as 
synchrotron radiation \citep{Rees_Meszaros1994,Tavani1996,Papathanassiou1996,Beniamini_etal_2018} or emissions originating from the photosphere, wherein subphotospheric dissipation \citep{Rees&Meszaros2005,Peer&Waxman2004,Peer&Waxman2005,Beloborodov2010} leads 
to spectral shapes that significantly deviate from a simple blackbody or that from a non-dissipative 
photosphere \citep{Pe'er2008,Beloborodov2011,Lundman2013}. Nevertheless, our study indicates that the emission detected from GRB131014A is unlikely to originate from these aforementioned two processes. Below we discuss the reasons supporting this inference.  

The distinguishing aspect of the GRB131014A spectrum lies in the occurrence of multiple breakpoints. As evident in Figure \ref{nuFnu}, the $\nu F_{\nu}$ plot 
exhibits several peaks denoted as $E_{peak1}$, $E_{peak2}$, and $E_{peak3}$. Notably, there is an additional break marked by 
the extended high-energy power law. The spectrum beyond this break deviates significantly from the spectral turnover observed after 
$E_{peak3}$. 

Furthermore, within the burst's bright region, it's observed that the low-energy power law index predominantly maintains values $> -0.5$ and, in certain instances, approaches $+1$. Such hard $\alpha$ values ($\alpha > -0.5$) are incompatible with optically thin synchrotron radiation, which has a theoretical “line of death” at $\alpha = -2/3$ or softer \citep{Preece1998,Burgess_etal_2015}. Moreover, the study by 
\citealt{Acuner_etal_2019} showed that synthetic spectra generated from non-dissipative photospheric emission models—when folded through the Fermi GBM response and fitted with empirical models like the Band function—typically yield $\alpha$ values in the range 
$-0.4$ to $0.0$. This aligns closely with our observed $\alpha$ values, suggesting a substantial thermal contribution at low energies. Furthermore, several GRBs exhibiting spectra fully consistent with a pure Planck (blackbody) function, including GRB 101219B 
\citep{Larsson2015}, GRB 100507 \citep{Ghirlanda2013}, and others observed by BATSE \citep{Ryde2004} have also been observed. In the time resolved spectral analysis, it is expected that some spectral smearing can be happen in the resultant time intervals and therefore the observed hard spectral slopes ($\alpha > -0.5$) are typically attributed to thermalisation processes. In some time bins likely minimally affected by 
spectral smearing—we observe even harder slopes approaching $\alpha \sim +1$, consistent with pure Planckian emission (as evident from the $\nu F_{\nu}$ plot, Figure \ref{nuFnu}b, where the asymptotic low energy part is entirely captured by the BB component). Both the scenarios reinforce the physical plausibility of a thermal component contributing significantly to the observed spectrum, particularly at low energies.  These results are also in agreement with earlier findings such as those by \citet{Guiriec_etal_2015}, as discussed in Section \ref{non_detection_BB}.

In terms of spectral slope transitions, there are four break points across the observed spectrum. This structure of multiple breaks invites comparison with theoretical synchrotron spectra, such as those described by \citealt{Sari1998}, where fast and slow 
cooling regimes are characterised by three spectral breaks. Since our analysis shows a hard low-energy slope (i.e., $\alpha > -2/3$) below $E_{peak1}$ - indicative of thermalisation - $E_{peak1}$ may correspond to the synchrotron self-absorption frequency. 
Under this interpretation, $E_{peak2}$ and $E_{peak3}$ could plausibly be associated with the cooling frequency $\nu_c$ and the characteristic synchrotron frequency $\nu_m$ in either fast or slow cooling regimes respectively. In that case, the emergence of the 
additional spectral break associated with the extended power-law component, particularly at higher energies, lacks a clear analog within standard synchrotron models.

Furthermore, the steep $\alpha$ values, approaching nearly $+1$, if 
associated with the spectral region below synchrotron absorption frequency, would necessitate an exceptionally high bulk Lorentz factor ($\Gamma \ge 1000$) of the 
outflow and also requires huge magnetic fields in order to position the absorption peak within the X-ray frequency range \citep{Granot_etal_2000, Lloyd_Petrosian2000}. Among the diverse modified versions of synchrotron emission models, 
Burgess et al. \citep{burgess_etal_2019NatAs} proposed a synchrotron model that incorporates time-dependent cooling of accelerated electrons. This model was demonstrated to successfully fit 
approximately $95\%$ of time-resolved peak GRB spectra, encompassing both soft and hard $\alpha$, surpassing even the $\alpha = -0.67$ threshold known as the "line of death" for synchrotron emission \citep{Preece1998}. However, within these notably successful fits illustrated in Figure 4 of 
Burgess et al. \citep{burgess_etal_2019NatAs}, the likelihood of achieving a successful fit to a spectrum characterised by $\alpha = +1$ is exceedingly low.

Alternatively, another model to explain the GRB spectrum is the sub-photospheric dissipation emission model.
The presence of multiple breakpoints within the spectrum negates the plausibility of continuous dissipation occurring from far below up to the photosphere \citep{Beloborodov2010,Giannios2012,Beloborodov2013}. When 
exploring localised sub-photospheric dissipation models, such as those discussed in \citep{Peer&Waxman2004, Peer&Waxman2005}, the studies by Ahlgren et al. \citep{Ahlgren_etal_2015,Ahlgren_etal_2019,Ahlgren_etal_2022yCat} wherein the model is directly tested with data, has revealed that for 
sub-photospheric dissipation occurring at moderate optical depths, the spectrum at high energies beyond the highest peak tends to exhibit a very steep profile 
(refer to the $\nu F_{\nu}$ plots depicted in Figure 1 in Ahlgren et al. \citep{Ahlgren_etal_2019}). However, in the case of GRB131014A, the high-energy spectrum doesn't showcase a cutoff; 
instead, it displays an extended power law characterised by a slope of approximately $-2.4$, persisting until energies around 
a few GeV. 

Therefore, upon evaluating the observational spectral characteristics of the burst in contrast to the expected behaviours of synchrotron and sub-photospheric dissipation 
models, the prevailing physical scenario appears to be consistent with optically thin inverse Compton scattering of the seed thermal photons originating from the photosphere.

\subsection{Proposed Physical Scenario: Optically thin Inverse Compton}  \label{ICS eqns} 
In the baryonic fireball model \citep{goodman,pzky_fireball} scenario, the prompt gamma ray emission is composed of the thermal emission that gets decoupled from the jet photosphere while the non-thermal emission is produced in the optically thin region above the photosphere.  
The kinetic energy of the outflow may get dissipated via processes like internal shock mechanism \citep{internal_shock1,internal_shock2,internal_shock3,internal_shock4}, collisional 
dissipation \citep{collisonal_dissipation}, slow heating via plasma instabilities or turbulences \citep{Ghisellini&Celotti1999,Pe'er2006} etc., in the optically thin region resulting in shocks wherein the electrons get accelerated to relativistic speeds. In the absence of strong 
magnetic fields, the electrons lose energy via upscattering of the soft thermal photons advected from the photosphere leading to the formation of a non-thermal spectrum. This scattering processes is referred to as the optically thin inverse Compton scattering (ICS). In modeling the spectrum of the burst GRB 131014A, we examine inverse Compton scattering involving only single order of scatterings within an optically thin medium (please refer section \ref{Highorder_scattering} for more details). This results in an ICS spectrum which conforms to the electron distribution's profile (refer to Appendix \ref{appendix2}).
The post-shock electron distribution in the outflow is considered to be encompassed of electrons in a thermal pool as well as those accelerated into a power-law tail, as given by
\begin{equation}
n_e(\gamma_e) = n_0 \left[ 
\frac{\gamma_e^2 \sqrt{1 - \frac{1}{\gamma_e^2}}}{\theta \, K_2\left(\frac{1}{\theta}\right)} e^{-\frac{\gamma_e}{\theta}} + \epsilon \left(\frac{\gamma_e}{\theta}\right)^{-\delta} \Theta \left(\frac{\gamma_e}{\gamma_{min}}\right)
\right]
\label{eqn1}
\end{equation}

where, $n_e$ is the number of electrons with Lorentz factor, $\gamma_e$; $n_0$ is the normalisation, $\theta$ = $\frac{kT}{m_e c^2}$ is the temperature of the thermal electrons in the units of electron's rest mass energy and the corresponding electron thermal Lorentz factor ($\gamma_{th}$) is $\theta$ + 1. $\gamma_{min}$ is the minimum electron Lorentz factor for the power law tail and is considered to be $\kappa$$\theta$ + 1 where $\kappa$ is an arbitrary constant,
$\epsilon$ is the normalisation of the power law, and $\delta$ is the power law index, $K_2$ is the modified Bessel function of second kind and $\Theta$(x) is the step function with $\Theta$(x) = 0 for x $<$ 1 and $\Theta$(x) = 1 for x $\ge$ 1. Unlike the ultra-relativistic thermal electron distribution($\theta$ $>>$ 1) used by \cite{Burgess2014a} and \citep{Baring_Braby2004}, we adopt the general form of the Maxwell-Jüttner distribution to represent the thermal component of the post-shock electron population, offering the flexibility of a more general thermal distribution. In the ultra-relativistic limit ($\theta >>$1), $\theta \to \gamma_{th}$ and $\gamma_{min} \to \kappa\gamma_{th}$ matching the post-shock electron distribution considered by \citealt{Baring_Braby2004} and \citealt{Burgess2014a}.   
In this paper, we consider $\kappa$ = 3 motivated from the studies on the simulations of particle acceleration at relativistic shocks which shows that the non-thermal population is directly derived from the superthermal tail of the thermal distribution \citep{Baring_1993,Baring_Braby2004}.
Furthermore, 
following the approach in \cite{Burgess2014a}, in order to have a minimised discontinuous transition between the thermal and non-thermal parts of the electron distribution, the value of $\epsilon$ is set to a small numerical value of 
\[
\epsilon =  \left[ 
\frac{\gamma_{min}^2 \sqrt{1 - \frac{1}{\gamma_{min}^2}}}{\theta \, K_2\left(\frac{1}{\theta}\right)} e^{-\frac{\gamma_{min}}{\theta}}
\right]
\label{eqn2}
\]




\noindent
Note that due to single-order scattering, the thermal pool of electrons is anticipated to generate a corresponding spectral peak, while the extended power law distribution of the electrons is expected to produce a power law within the observed spectrum at higher energies.

The degree of Comptonisation in a medium is defined by the Compton $y$ parameter \citep{rybicki} which is given by the 
product of the average number of scatterings and the average fractional photon energy change per scattering. If $y$ $\gtrsim$ 1, the incident photon energy and the overall 
spectrum will undergo significant changes. If $y$ $\ll$ 1,  the incident photon energy and spectrum minimal change is expected.
Neglecting the down scattering of photons, the up-scattered photon energy ($x_f$) can be expressed in terms of the Compton $y$ parameter and the photon energy ($x_0$) before scattering as  \citep{gabriele}
\begin{equation}
   x_f = x_0 e^y \label{eqn3}
\end{equation}
The mean amplification $A$ of the photon energy at each scattering is given by, 
\begin{equation}
A = \frac{x_f}{x_0}  
\label{A}
\end{equation}
The equations \eqref{eqn3} and \eqref{A}, yields the Compton $y$ parameter in terms of the mean amplification factor as follows,
\begin{equation}
   y = ln(A) \label{eqn5}
\end{equation}
Furthermore, in the optically thin relativistic medium, the average number of scatterings is equivalent to the optical depth at the site, $\tau$ and using the definition of Compton $y$ parameter \citep{Rybicki&Lightman1986,gabriele}, the optical depth at the dissipation site can be estimated as  
\begin{equation}
   \tau = \frac{y}{A - 1} \label{eqn6}
\end{equation}
Knowing the optical depth $\tau$ at the dissipation site further allows to estimate the dissipation radius as follows 
\begin{equation}
   R_d = \frac{R_{ph}}{\tau} \label{eqn7}
\end{equation}
where $R_{ph}$ is the photospheric radius.\\

The average energy of the up-scattered photon can be expressed in terms of the average electron's Lorentz factor in thermal part of the electron distribution, $\langle \gamma_e \rangle$, \citep{longair, rybicki}, as follows 
\begin{equation}
   x_f = \left (\frac{4}{3} \langle\gamma_{e,th}^2 \rangle \langle \beta_{e,th}^2 \rangle \right) x_0 \label{avg_energy}
\end{equation}
where, $\langle \beta_{e,th}^2 \rangle$ 
= $(1 - \frac{1}{\langle \gamma_{e,th}^2 \rangle})$ = $\frac{\langle v_e^2 \rangle}{c^2}$, $\langle v_e^2 \rangle$ is the average of the square of electron's velocity in the thermal part of the electron distribution and $c$ is the velocity of light. 
Using equations \eqref{A} and \eqref{avg_energy}, the average electron Lorentz factor can be expressed as follows:
\begin{equation}
    \langle \gamma_{e,th}^2 \rangle = \frac{3}{4}A + 1  
    \label{gamma_e_A}
\end{equation}

At asymptotic higher energies, the ICS spectrum generated by the power-law distribution of electrons is approximately described as follows \citep{atoyan} \footnote{The power law dependence at higher energies also includes a logarithmic factor of $ln$($\frac{x_f}{m_e}\frac{x_0}{m_e}$), where $m_e$ is the mass of the electron. However, we note that the contribution of this factor remains nearly constant at the high energy asymptotic limits of the observed spectrum.}
\begin{equation}
    C_{asym} \propto x_f^{-(\delta + 1)}
    \label{delta}
\end{equation}
where $C_{asym}$ is the counts flux of ICS photons at the asymptotic higher energies.

\subsubsection{Estimating Physical Parameters from Observables}

In the proposed physical scenario, we recognize ($E_{\text{peak2}}$ + $E_{\text{peak2}}$)/2 as the peak energy resulting from the first-order inverse Compton scattering of the blackbody (BB) component ($E_{peak1}$) by electrons in the thermal portion of the electron distribution. The amplification factor $A$, is thereby determined as follows
\begin{equation}
   A = \frac{E_{peak2} + E_{peak3}}{2E_{peak1}} \label{A-est}
\end{equation}
The higher energy spectral peak part of the spectrum is produced by the inverse Compton scattering off the thermal electrons with Lorentz factors ranging between $\gamma_e$ = 1 to 2$\gamma_{min}$.
Thus, the average electron Lorentz factor of the thermal part of the electron distribution, $\langle \gamma_{e,th}^2 \rangle $ applicable in equation \eqref{gamma_e_A} is estimated as follows
\begin{equation}
   \langle \gamma_{e,th}^2 \rangle\ = \frac{\int_{1}^{2\gamma_{min}}\gamma_e^2 n_e (\gamma_e)\,d\gamma_e }{\int_{1}^{2\gamma_{min}} n_e (\gamma_e)\,d\gamma_e }  \label{eqn12}
\end{equation}
Given the expression \eqref{delta} for the asymptotic high energy part of the ICS spectrum, using the observable of the high energy spectral index, $\beta$ allows to estimate the electron power law index $\delta$ as follows \eqref{delta}
\begin{equation}
   \delta = -\beta - 1  \label{eqn14}
\end{equation}

Using equations \eqref{gamma_e_A}, \eqref{A-est}, \eqref{eqn12} and \eqref{eqn14}, $\gamma_{min}$ is estimated numerically using the observed parameters $E_{peak1}$,$E_{peak2}$ and $E_{peak3}$.

The normalisation, $n_0$ of the electron distribution can be estimated using the estimate of total number of electron in the shock region,
\begin{equation}
   N_{e} = \frac{L_{obs}}{\Gamma m_p c^2 }\times t_{obs}
   \label{Nel_L}
\end{equation}
\noindent
where $L_{obs}$ is the observed burst luminosity, $m_p$ is the mass of proton, $t_{obs}$ is the observed time and $\Gamma$ is the bulk Lorentz factor. Here, $N_e = \dot{N_e} \times t_{obs}$, where the rate of change of particle number can be expressed as , $\dot{N_e} = \frac{L_{obs}}{\Gamma m_p c^2 }$.
\\
The total number of electrons in the shock region can also be evaluated by integrating the electron distribution function given by equation \eqref{eqn1} for all possible values of its Lorentz factor,
\begin{equation}
   N_{e} = \int_{1}^{\infty}n_e (\gamma_e)\,d\gamma_e  \label{eqn15}
\end{equation}
leading to the following
\begin{equation}
   N_{e} = n_0\left[ \int_{1}^{\infty}\frac{\gamma_{e}^2 \sqrt{1 - \frac{1}{\gamma_{e}^2}}}{\theta \, K_2\left(\frac{1}{\theta}\right)} e^{-\frac{\gamma_{e}}{\theta}}\,d\gamma_e + \epsilon \int_{\gamma_{min}}^{\infty}\left(\frac{\gamma_e}{\theta}\right)^{-\delta}d\gamma_e\right]  \label{Ne_eldist}
\end{equation}
Solving equations \eqref{Nel_L} and \eqref{Ne_eldist}, $n_0$ can be obtained in terms of $\gamma_{min}$, $\delta$, $L$ and the outflow parameters ($R_d$ and $\Gamma$, see section \ref{sec_outflow}) which in turn are estimated using the spectral observables.

The number of electrons in the power law tail of the electron distribution is given as follows
\begin{equation}
   N_{e,pow} = n_0  \epsilon \int_{\gamma_{min}}^{\infty}\left(\frac{\gamma_e}{\theta}\right)^{-\delta}d\gamma_e    
   \label{Nel_pow}
\end{equation}

\subsection{Outflow parameters} 
\label{sec_outflow}
In the physical model interpretation of the burst spectrum, the blackbody component identified in the best fit spectral model 
is considered as the thermal emission advected from the photosphere of the jet. 
Associating the observed BB component to the photospheric emission allows to estimate the jet outflow parameters such as the Lorentz factor ($\Gamma$) at the photosphere; the nozzle radius ($R_0$): the radius from where the jet starts to expand freely; the saturation radius ($R_s$): the radius where the internal energy density of the jet becomes equal to the kinetic energy density of the jet; and the photospheric radius ($R_{ph}$): the radius at which the photons get decoupled from the plasma, using the methodology given in \citep{asaf_2007,Iyyani2013,Iyyani2016}.  For an assumed redshift $z = 2.12$, an average value for the GRBs with LAT data and a radiative efficiency $1/Y = 0.65$ \citep{racusin_2011}, the average values of the outflow parameters are as follows: $\Gamma = 400 \pm 6$, $R_0$ = $(2.4 \pm 1.1 )\times 10^8$ cm, $R_s$ = $(8.4 \pm 0.8)\times 10^{10}$ cm and $R_{ph}$ = $(1.8 \pm 0.3) \times 10^{13}$ cm. 
 
\begin{figure*}[!ht]
    \centering
    \includegraphics[width=1.05\linewidth]{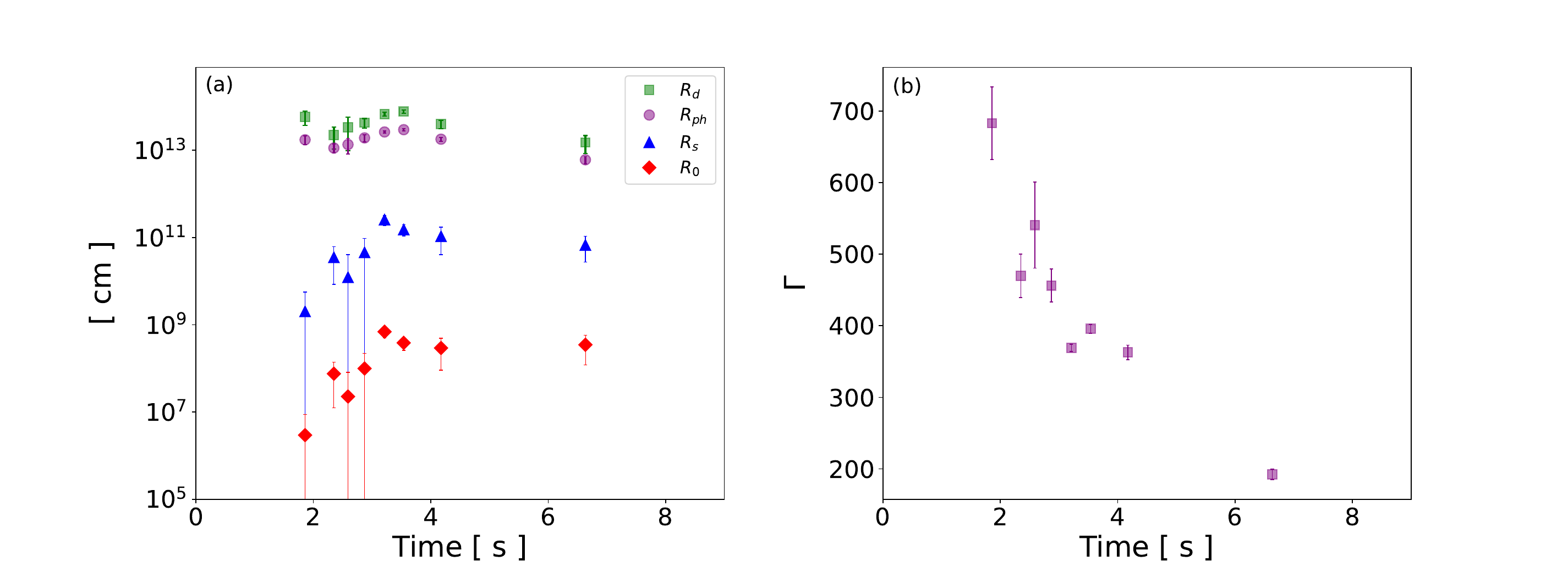}
    \caption{The evolution of the jet outflow parameters including (a) nozzle radius, $R_0$ (red diamond); saturation radius, $R_s$ (blue triangle); photopsheric radius, $R_{ph}$ (magenta circle), dissipation radius, $R_d$ (green square) and (b) the bulk Lorentz factor, $\Gamma$ with time are shown.}    
    \label{fig:5}
\end{figure*}

The temporal evolution of the outflow parameters is shown in Figure \ref{fig:5}, except for the first six bins up to 1.49 s, as the BB component is not constrained during this period.
Among the radial parameters, the nozzle radius, $R_0$ is found to increase with time from nearly close to 
the central engine of around several $10^6\, \rm cm$ to a peak value of around $10^9 \, \rm cm$ and later 
remains nearly steady with time (Figure \ref{fig:5}a). This temporal behaviour is consistent with that observed in previous studies \citep{Iyyani2013,iyyani_2015,Iyyani2016}. 
A corresponding similar trend is observed in $R_s$ as well. The photospheric radius is found to be nearly steady with only mild variations around $10^{13}$ cm (Figure \ref{fig:5}a). 

The Lorentz factor ($\Gamma$) of the outflow shows predominantly a monotonous decrease from an initial value of nearly $700$ to around $200$ across the total burst duration as evident in Figure \ref{fig:5}b. It is noteworthy that despite multiple emission pulses during the total burst duration, on average, the Lorentz factor of the outflow keeps decreasing with time. 

\subsection{Inverse Compton Characteristics}\label{ICS3}
The estimates of the physical parameters that characterize the optically thin inverse Compton 
scattering using the equations mentioned in subsection \ref{ICS eqns} are discussed here. The average estimates for the physical parameters are mentioned along with their temporal behaviours. 

\begin{figure*}[!ht]
    \centering
    \includegraphics[width=1.05\linewidth]{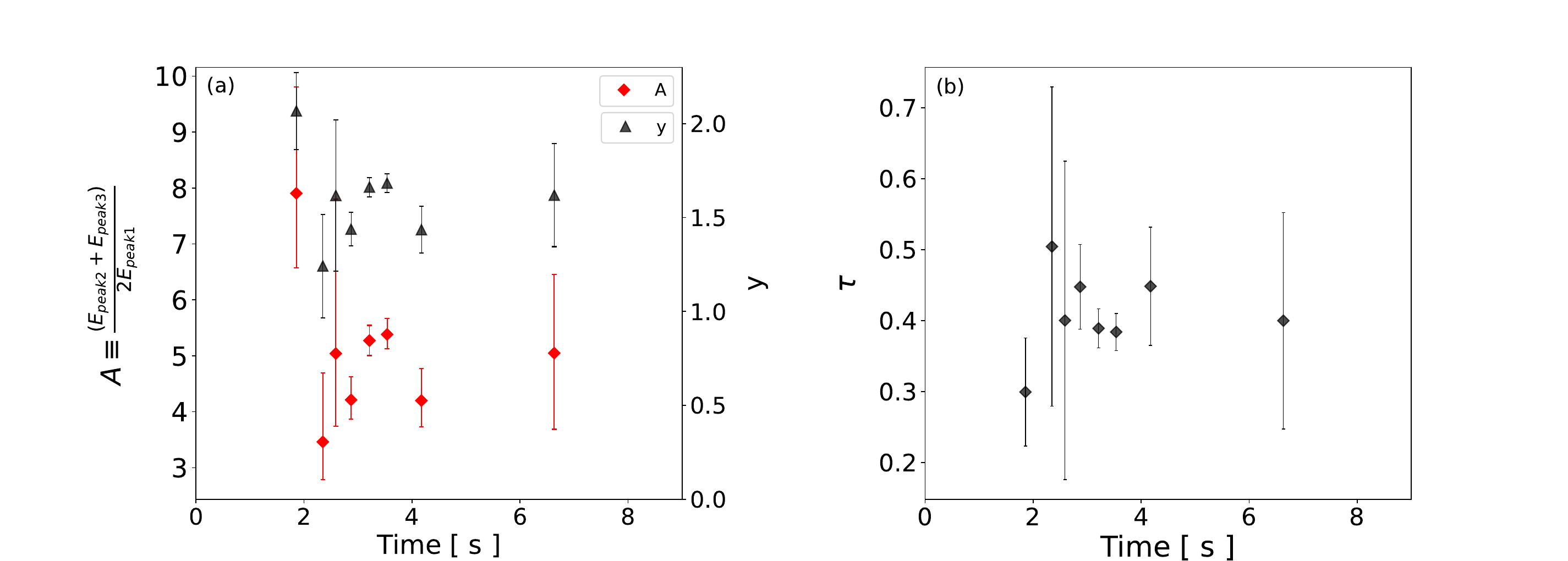}
    \caption{The temporal evolution of the microphysical parameters of the inverse Compton scattering such as (a) amplification parameter, $A$ (red diamond), Compton $y$ parameter (black triangle), and (b) optical depth $\tau$ (black diamond) at the dissipation site are shown.}
    \label{fig:6}
\end{figure*}

The mean amplification factor, $A$ and the Compton $y$ parameter are found to be $ 5.06 \pm 0.20$ and $ 1.60 \pm 0.08$ respectively. The value of $y \gtrsim 1$ suggests a substantial spectral change due to the inverse Compton scattering process. The temporal variation of these parameters is shown in Figure \ref{fig:6}a. 
The optical depth ($\tau$) at which the dissipation shocks are generated is found to be $0.41 \pm 0.05$ which indicates that the dissipation is taking place in the optically thin region of the jet (Figure \ref{fig:6}b). This further gives the estimate of
the dissipation radius which is found to lie closely above the photosphere, within the range of $10^{14}$ cm as shown in Figure \ref{fig:5}a. We note that no significant trend is observed in the temporal evolution of these parameters during the burst duration. 

\begin{figure*}[!ht]
    \centering
    \includegraphics[width=1\linewidth]{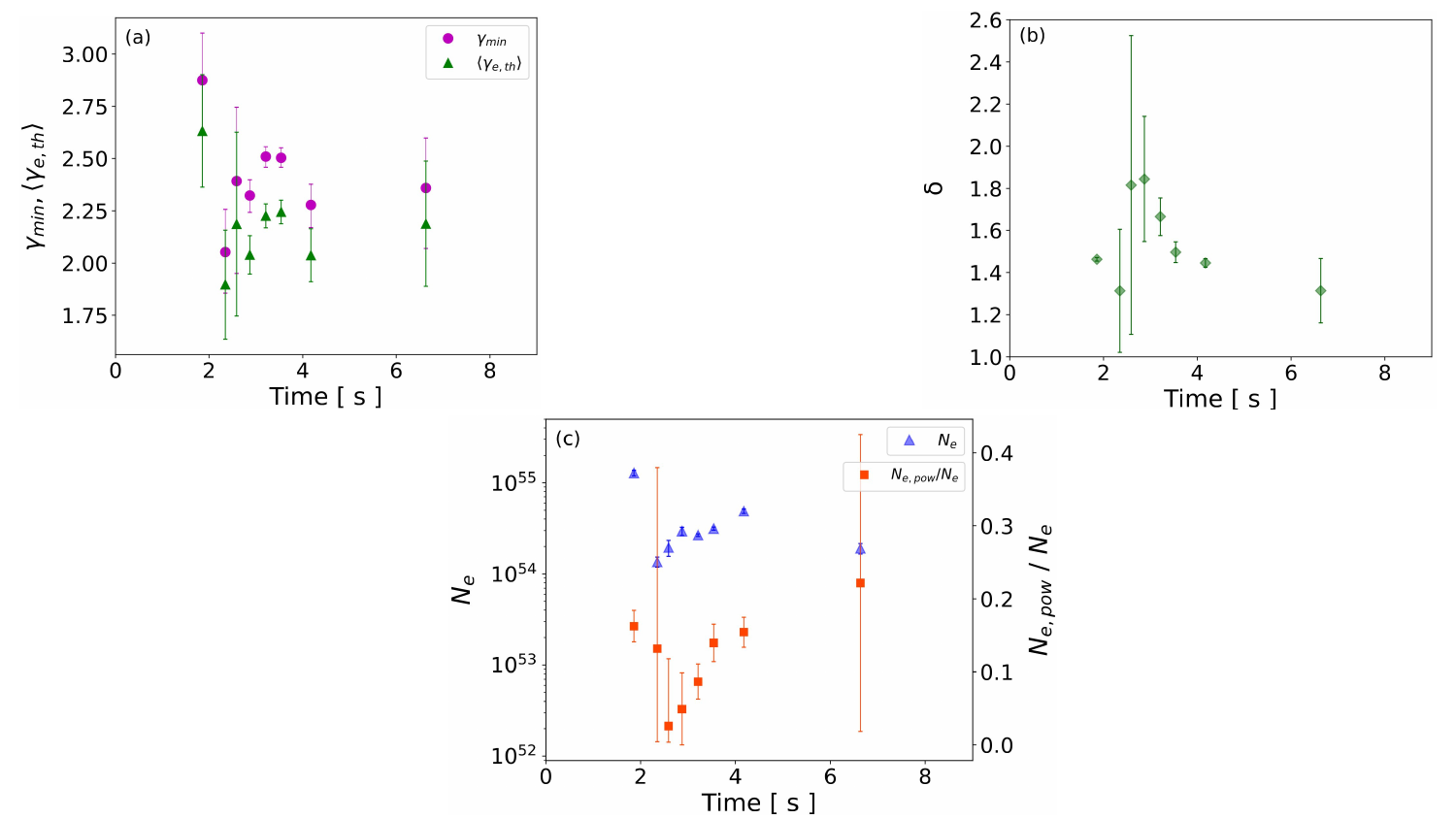}
    \caption{The temporal evolution of the parameters of the electron distribution at the dissipation site: (a) minimum electron Lorentz factor $\gamma_{min}$ (magenta circles), average electron Lorentz factor of the thermal part of the electron distribution, $\langle \gamma_{e,th} \rangle$  (green triangles), (b) power law index, $\delta$ (green diamond) and (c) total number of electrons, $N_e$ (blue triangle) along with the ratio of electrons in the power law with respect to the total number of electrons ($N_{e,pow}/N_e$, red square) are shown.} 
    \label{fig:7}
\end{figure*}

The estimates of the parameters characterising the electron distribution presented in section \ref{ICS eqns} are discussed below.  
The $\gamma_{min}$, $ \langle \gamma_{e,th} \rangle$, $\delta$ and $n_0$ are found to be $2.51 \pm 0.05$, $2.22 \pm 0.06$, $1.66 \pm 0.08$ and $ (1.11 \pm 0.14) \times 10^{54}$ respectively. The temporal evolution of these parameters 
are shown in Figure \ref{fig:7}a and \ref{fig:7}b respectively.
The total number of electrons, $N_{e}$ is found to be around $ 4 \times 10^{54}$, while the ratio of the number of electrons in the power law tail of the electron distribution to the total number of electrons, $N_{e,pow}$/$N_{e}$ is around $10\%$ on average which is consistent with the particle in cell (PIC) simulation studies \citep{Spitkovsky2008}.
The temporal variation of the parameters is shown in \ref{fig:7}c. The total number of electrons is found to increase by an order of magnitude to around $10^{55}$ during the initial pulses of the burst, however, later it decreases to around $10^{54}$ (Figure \ref{fig:7}c).

\subsection{Time-integrated spectrum - Results}\label{time_int_result}
The physical parameters, including the characteristic ICS parameters and the outflow parameters, estimated from the time-integrated spectrum of duration $8.6 \, \rm s$, using the methodologies described in sections \ref{ICS eqns} and \ref{sec_outflow} respectively, are reported here. The total observed luminosity, $L_{obs,int}$ is found to be 3.1 $\times 10^{54}$ erg/s. The outflow parameters are estimated to be: $R_{0,int}$ = $(2.9 \pm 0.5) \times 10^8$ cm, $R_{s,int}$ = $(7.1 \pm 1.2) \times 10^9$ cm, $R_{ph,int}$ = $(1.6 \pm 0.1) \times 10^{13}$ cm, $R_{d,int}$ = $(3.5 \pm 1.2) \times 10^{13}$ cm and $\Gamma_{int}$ = $434 \pm 10$. The parameters that characterize the optically thin inverse Compton scattering are as follows: $A_{int}$ = 6.20 $\pm$ 1.74, $y_{int}$ = 1.83 $\pm$ 0.28, $\tau_{int}$ = 0.35 $\pm$ 0.13, $\gamma_{min,int}$ = 2.75 $\pm$ 0.56, $\delta_{int}$ = 1.39 $\pm$ 0.02. The total number of the electrons estimated for the time-integrated spectrum is found to be $N_{e,int} = (4.10 \pm 1.44) \times 10^{55}$. These estimates represent the average behavior of the burst dynamics and are therefore used for calculations across various subsections of the Discussion section \ref{discussion}.
\\

\section{Discussion}  \label{discussion}

\subsection{Multiple breaks in GRB spectrum}
A primary observational finding in the spectral analysis of {\it Fermi}-detected GRBs indicates that among the 
brightest instances, the time-resolved GRB spectra significantly differ from a singular Band function \citep{Ackermann2013}. These variations involve multiple additional components, such as a blackbody function at lower 
energies \citep{Guiriec2011, Axelsson2012, Burgess2014a, Iyyani_etal_2015}, an extra power law extending to higher energies \citep{Abdo2009_090902B,Ryde2010,Ryde2011,Iyyani_etal_2015,Sharma_etal_2019}, a power law with an 
exponential cutoff \citep{Ackermann2011} or presence of multiple breaks demonstrated by models such as 2 smoothly broken power laws \citep{Ravasio_etal_2018,Ravasio_etal_2019A,Toffano_etal_2021}. Some cases even show multiplicative components, like an exponential cutoff at higher energies \citep{Ackermann2011, Vianello2018, Sharma_etal_2019}.

GRB131014A aligns with these findings, displaying a spectrum that doesn't conform to a simple, conventional Band function alone. The spectrum of GRB131014A exhibits distinctive features, including a thermal component at 
lower energies, as well as two spectral breaks identified at energies $E_{peak2}$ and $E_{peak3}$. Moreover, the trend of the turnover of 
the spectrum beyond $E_{peak3}$ is obscured by the extension of the high-energy power law, resulting in an additional 
spectral break at higher energies, indicating the onset of the extended power law while there is no need of power-law component extension to the lower energies as evidenced by the presence of hard $\alpha$ values 
($> -0.5$). This high-energy break and extended power law persist throughout the entire burst, from its onset to its conclusion. For 
typical parameters, such as a Lorentz factor \( \Gamma = 100 \) and an interstellar medium (ISM) located at \( 10^{16} \, \mathrm{cm} \), the expected delay in the onset of afterglow emission would be 
$t_{\text{delay}} = \frac{R}{2\Gamma^2 c} \sim 17 \, \rm{s}$.
Observing this feature from the very beginning, with a delay of less than a second, would require $ \Gamma$ to be in the range of thousand.
Additionally, the power-law index, $\beta$, of this high-energy component is consistently found to be $ < -2.3 $ (Figure \ref{spec_param}a), aligning with the typical behavior of high-energy components observed in the prompt emission of GRBs \citep{GBMcatalog2014}. In contrast, emerging afterglow components exhibit flatter or harder power-law indices ($\beta \ge -2$; \citealt{Ackermann_etal_2010,Ackermann_etal_2013,Ajello_etal_2018}). Thus, the extended power law does not appear to represent a distinct spectral component originating from a separate emission site, such as the afterglow. 

While multi-break spectral shapes have been reported previously, such a distinct spectral structure in a GRB spectrum is presented here for the first time. Furthermore, we note that the ratio between the highest spectral peak ($E_{peak3}$) and the lowest spectral peak ($E_{peak1}$) is approximately about an order of magnitude. 
The complexity of this spectral shape poses challenges for modeling with simple empirical functions such as the Band function, cutoff power law, or smoothly broken power law. Consequently, multiple empirical functions, including two Band functions and a blackbody, were employed in an additive fashion in order to capture its features.

Furthermore, we note that GRB131014A was previously examined by Guiriec et al. \citep{Guiriec_etal_2015}, who focused solely on GBM data, excluding LLE and LAT emissions. Consequently, they may have overlooked the high-energy spectral break and extended power-law identified in our study. 

\subsection{Non-detection of BB in initial bins}
\label{non_detection_BB}

\citealt{Guiriec_etal_2015} observed a robust thermal component in the spectrum of GRB131014A while conducting a 
detailed time-resolved spectral analysis with bin widths ranging from 2 ms to 94 ms \citep{Guiriec_etal_2010}. This 
analysis revealed initially hard low-energy spectral slopes ranging from $0$ to $+1$ during the burst's early stages. 
In contrast, our study employs Bayesian Block binning for time-resolved analysis, resulting in broader time intervals and relatively softer $\alpha$ values ($\ge -0.5$) for the 
considered model. \citealt{Acuner_etal_2019} demonstrated that synthetic spectra generated by convolving 
the theoretical model of non-dissipative photospheric emission ($\sim $ thermal component) with {\it Fermi} GBM's response yielded 
$\alpha$ values ranging between $-0.4$ and $0.0$ when modeled using empirical functions like the Band function. 
Hence, our observation of $\alpha > -0.5$ strongly suggests a notable thermal contribution at lower energies 
within the burst spectrum, aligning with the findings of \citealt{Guiriec_etal_2015} where notably, their Band-only fits 
yield hard low-energy indices (predominantly $+0.5 \ge \alpha \ge -0.5$) throughout the burst. However, in their multi-component fits, $\alpha$ is fixed at $-0.7$, possibly due to 
limited constraints in fine time bins, making direct comparison with our freely fitted low energy spectrum unfeasible.
Additionally, we 
observe that the thermal peak values reported by \citealt{Guiriec_etal_2015} during the initial time bins until 1.49 s range from approximately 100 keV to 200 keV (refer to Figure 2c in \citealt{Guiriec_etal_2015}), which 
corresponds well with the $E_{peak2}$ values identified in our analysis (refer to Figure \ref{spec_param}b). Thus, it is worth noting that 
the thermal component of the best-fit model discussed in this current study differs from that referenced in Guiriec et al.'s work. Therefore, thermal fluxes and related outflow parameters (for $z = 1.55$) presented in their work are not directly comparable. Given the differences in data, time binning, and spectral modeling, a strict comparison is not possible. Nonetheless, both studies converge on the presence of thermalisation at low energies, as indicated by hard low energy spectral slopes.
\\

In our physical model interpretation, we have considered the GRB131014A to possess a baryon-dominated jet, which is typically expected to produce prominent thermal detections, in contrast to Poynting flux-dominated outflows, where the photospheric emission is highly suppressed \citep{Zhang_Pe'er2009}. However, our analysis fails to constrain the thermal component in the initial bins until 1.49 s.  
On extrapolating the observed trend of the ratio of thermal flux to the observed total flux (Figure \ref{spec_param}c) in to the earlier time bins indicates that the expected 
thermal flux is much less than $1\%$. In addition, the trend of the thermal peak ($E_{peak1}$) on extrapolating to times before 1.49s also indicates values less than 40 keV which corresponds to $kT \ll 10 \, \rm keV$. As the thermal component approaches lower fluxes as well as the edge of the observation window of the {\it Fermi} GBM makes it technically and statistically challenging to determine the component in the early time bins.    

\subsection{Microphysics of shock and electron cooling}
\label{microphysics}

\subsubsection{Pressure Anisotropy and Magnetic Obliquity in Shock Region}

The electron power-law index ($\delta$) is generally expected to be around 2 to 2.2 for diffusive 
acceleration at the parallel, non-relativistic or ultra-relativistic shocks under isotropic pressure conditions \citep{kirk_1987,kirk_2000}.  However, \citealt{Shen_etal_2006} demonstrated that the electron power-law indices, derived from the high
energy spectral power-law index ($\beta$), span a broader range of approximately 1.6 to 4 across various astrophysical sources, including GRB prompt emissions, X-ray afterglows, blazars, and 
pulsar nebulae. Furthermore, astrophysical environments with relativistic shocks, pressure anisotropy and magnetic fields being oblique to the shock normal can be expected.

Collisionless shocks in an optically thin region of the outflows form via dissipation mechanisms like internal shocks, typically resulting in mildly 
relativistic shock velocities. As these shocks propagate, even small upstream magnetic field angles relative to the 
shock normal are significantly altered by Lorentz transformation into the downstream frame, leading to 
stronger and more oblique magnetic fields. Consequently, relativistic shocks require oblique magnetohydrodynamic (MHD) jump conditions for accurate modeling.
Most studies on relativistic shock jump conditions assume that incoming particles do not affect the shock structure 
and that pressures remain isotropic \citep{Kirk&Webb1988}. However, pressure anisotropy naturally arises in relativistic shocks, as particles struggle to diffuse upstream against the relativistic 
flow, leading to anisotropic ion and electron distributions. This anisotropy is pronounced in the shock transition layer and persists even at ultrarelativistic energies. \citealt{Double_2004} numerical solutions 
reveal that relatively small departures from pressure isotropy can produce significant changes in the shock compression ratio, $r_c$, at all shock Lorentz factors. 
Thus, the pressure anisotropy and magnetic obliquity at the shocks may impact models of particle acceleration such as diffusive shocks acceleration mechanism wherein a 
certain fraction of electrons from the thermal pool gets accelerated to very high energies, producing a high energy power law tail.  
The values of the power law index, $\delta$, of the electron distribution \ref{eqn1}, thereby, reveals some crucial characteristics of the shock and the background medium.

The pressure anisotropy is characterised by:  

\[
\alpha_p = \frac{P_{\perp}}{P_{\parallel}}
\]
\noindent
where $P_{\perp}$ and $P_{\parallel}$ are the pressures perpendicular and parallel to the magnetic field, respectively. Simulation studies done by \citealt{Double_2004} have shown that when $\alpha_p < 1$, the compression ratio $r_c$—defined as the ratio of downstream to 
upstream plasma densities—can exceed 3, the expected value for relativistic shocks with isotropic pressure. This increase in $r_c$ leads 
to a softening of the electron power-law index ($\delta$) due to pressure anisotropy, following the relation:

\[
\delta = \frac{r_c + 2}{r_c - 1}
\]
\noindent
This relation is valid for non-relativistic shocks, and also similar values are obtained for ultra-relativistic case, as numerical simulations indicate that for $r_c=3$, the electron power-law index is approximately $\delta \sim 2.3$ \citep{Baring_2006}.  

\citealt{Double_2004} further noted that in ultra-relativistic shocks, the obliquity of the 
magnetic field does not significantly affect the variation of $r_c$ with $\alpha_p$. However, in mildly relativistic shocks, changes in the 
electron power-law index depend on both $r_c$ and the obliquity of the magnetic field relative to the shock normal. As shown in 
\citealt{Summerlin2012}, for a fixed compression ratio of approximately $3.7$, the electron power-law index $\delta$ becomes significantly flatter 
($\delta < 2$) in the magnetic field obliquity range of $20^\circ - 40^\circ$, provided the diffusion parameter satisfies the physically meaningful condition:

\[
\frac{\lambda}{r_g} \gg 1
\]
\noindent
where $\lambda$ is the mean free path of diffusion and $r_g$ is the particle gyroradius.  

Thus, the observation of softer ($\delta$) values ($< 2.2$, Figure \ref{fig:7}b) suggests the presence of both pressure anisotropy and significant magnetic field obliquity within ($20^\circ - 40^\circ$) of the shock normal, leading to more efficient particle acceleration to higher energies.
\subsubsection{Average Electron Lorentz Factor and Cooling Timescale Analysis}
\label{average_electron}
The radiation efficiency of GRBs can be expressed as the product of three key fractions: the fraction of 
the total burst energy that is dissipated ($f_d$), the fraction of the dissipated energy transferred to electrons ($f_e$), and the fraction of electron energy that is radiated ($f_{er}$). Based on 
\citealt{racusin_2011}, GRBs similar to GRB 131014A exhibit a radiation efficiency of approximately $0.65$, which suggests that the fraction\footnote{This value is derived under the assumption that the three factors mentioned above 
contribute comparably (each fraction's value is $\approx$ 0.87) to the overall radiation efficiency. If one of the fractions takes values below 0.65, at least one of the other 
fractions would have to exceed 1, which is physically unrealistic. The obtained $\epsilon_e$ value aligns with the 
proposed scenario, where a significant fraction of the burst energy is transferred to electrons via dissipation rather than to magnetic fields.} of the total burst energy carried 
by electrons ($\epsilon_e = f_d \times f_e$) to be around $0.76$.  

The total energy of the electrons in the lab frame is given as
\begin{equation}
E_{el} = \epsilon_e E_{iso} =  \Gamma N_e \, \langle\gamma_e\rangle \, m_ec^2
\label{eqn_avg_egamma}
\end{equation}
where $E_{iso} = 2.7 \times 10^{55} \, \rm erg$, $\Gamma = 434$ and $N_e = 4.1 \times 10^{55}$ for the time-integrated spectrum.  
The average electron Lorentz factor is estimated to be of the 
order of a few thousand $\langle \gamma_e \rangle \sim 1400$. Consequently, the maximum Lorentz factor of electrons can be estimated as $\gamma_{\max} \sim$ a few $ 10^4 $ using the
equation~\ref{eqn1}, where the values of the electron distribution parameters estimated for the time-integrated spectrum, derived in Section~\ref{time_int_result}, are applied.

We, however note that the highest observed photon energy is only $1.8\, \rm GeV$ despite the average electron energy of the overall electron 
distribution is around $10^3$. The suppression of observation of very high-energy GeV photons could be attributed to factors such as 
(a) pair production constraints at the dissipation site:  If the compactness parameter at the dissipation site exceeds 1, high-
energy photons produced in the co-moving frame may undergo pair production, leading to their 
suppression in observations (see Section \ref{Highorder_scattering} for more details); (b) for distant GRBs, GeV photons also have a 
significant probability of interacting with the extragalactic background light (EBL), resulting in pair production and subsequent attenuation.

We note that, for this burst, the obtained $\gamma_{\text{min}}$ values are mildly relativistic (Figure \ref{fig:7}), as may be expected from dissipation mechanisms 
such as internal shocks\footnote{In internal shock scenarios, the minimum Lorentz factor of accelerated electrons can be estimated as $\gamma_{\rm min} \approx \epsilon_e \epsilon_d 
(\Gamma_{\rm rel} - 1) m_p / (\zeta_e m_e)$, where $\epsilon_d$ is the fraction of kinetic energy dissipated, $\epsilon_e$ is the fraction of dissipated energy transferred to electrons, 
$\zeta_e$ is the fraction of electrons accelerated, and $\Gamma_{\rm rel}$ is the relative Lorentz factor between colliding shells. For shells with comparable Lorentz factors 
(Figure \ref{fig:5}b), using typical values $\epsilon_d \sim 
0.02$, $\epsilon_e \sim 0.1$, $\zeta_e \sim 0.1$, and $\Gamma_{\rm rel} \sim 1.06$, we find $\gamma_{\rm min} \sim 2.2$, consistent with mildly relativistic internal shocks 
\citep{Kobayashi_etal_1997,Daigne&Mochkovitch1998,Zhang_etal_2007,Spitkovsky2008,racusin_2011}.}. Such shocks, arising from impulsive energy injections, are expected during the prompt 
emission phase of gamma-ray bursts, where colliding shells typically possess comparable Lorentz factors \citep{Daigne&Mochkovitch1998,Kobayashi_etal_1997}. The overall 
average electron Lorentz factor, $\langle \gamma_e \rangle$, when accounting for the full electron distribution, can still reach values of a few thousand in these scenarios 
\citep{Daigne&Mochkovitch1998}. In contrast, external forward shock synchrotron models generally require a much larger minimum electron Lorentz factor, $\gamma_{\rm min} \sim (m_p/m_e) \Gamma \sim 10^4$–$10^5$ \citep{Sari1998}. 

We emphasise that the $\gamma_{\rm min}$ values derived in this work are only mildly relativistic, unlike most existing 
estimates, often several thousand or higher, which are typically inferred either from synchrotron modelling of the prompt emission (e.g., \citealt{Burgess2014a,Iyyani2016}) or 
from afterglow studies based on external shock dynamics where synchrotron emission is assumed to dominate (e.g., \citealt{Sari1998,Racusin2011}). By contrast, the current analysis characterises the electron distribution and the 
electron $\gamma_{\rm min}$ values within a framework where inverse Compton scattering is considered as the prompt 
radiation mechanism.

Furthermore, it is worth noting that internal shock dissipation scenarios generally require relatively modest dissipation 
efficiencies \citep{Kobayashi_etal_1997}. Although the redshift of the burst is not known, its high fluence implies that the dissipation efficiency would need to be of the order of tens of percent; 
otherwise, unrealistically large kinetic energies of the jet would be required. This suggests that internal shocks alone are unlikely 
to account for the entire dissipation process.

In addition, several alternative dissipation mechanisms have been proposed in the literature that can produce heated Maxwellian 
distributions of mildly relativistic electrons. Of particular interest is the slow-heating scenario \citep{Ghisellini&Celotti1999,Pe'er2006}, in which, just above the photosphere ($\tau_{\gamma e} \lesssim 1$), the jet’s kinetic energy 
is gradually dissipated through plasma instabilities and turbulence. This maintains electron energisation over timescales comparable to the dynamical time, in contrast to the impulsive heating typical of 
standard internal shocks. While direct quantitative estimates of the dissipation efficiency for such cases are scarce, the extended dissipation period suggests that efficiencies of several tens of percent may be achievable. In this dissipation scenario, the electron population forms a Maxwellian distribution with thermal 
Lorentz factors of a few, following $(\gamma_{e,th} \beta_{e,th})^{2} \approx \tau_{\gamma e}^{-1}$. These electrons Comptonise the advected thermal photons, yielding a spectrum 
consistent with prompt GRB observations, including a flat and broadened $\nu F_\nu$ component above the thermal peak. For our case, this mechanism is particularly relevant, as we estimate 
dissipation occurring at radii $\sim \rm few \times 10^{13} \, \rm cm$ with optical depths $\tau \sim 0.35$ (see Section~4.5), implying $\gamma_{e,th} \beta_{e,th} \sim 3$. 

It is also reasonable to expect that the actual dissipation process in GRB jets may involve a combination of mechanisms. For example, a hybrid scenario may occur just above the photosphere, where weak 
internal shocks trigger turbulence driven slow heating of electrons into a Maxwellian distribution with $\gamma_{e,th} \sim 2-5$, while a small fraction of electrons are accelerated into a power-law tail 
beginning at similarly low $\gamma_{\rm min}$. Such electron distribution profiles are supported by PIC simulations (e.g., \citealt{Spitkovsky2008}).

In this work, however, we do not aim to advocate for any one specific dissipation mechanism, as the nature of kinetic energy 
dissipation in GRBs remains an open question. Rather, we interpret the best fit spectral model to the observed data using optically thin inverse Compton scattering and infer the electron distribution parameters that best match the 
data. The low $\gamma_{\rm min}$ values we report are physically plausible under certain dissipation frameworks applicable to the prompt phase. Furthermore, the inferred electron distribution 
characteristics here may serve as a basis for the community to further investigate possible dissipation mechanisms and particle acceleration microphysics in GRB jets.
\\

For the ICS to be the dominant emission process over the synchrotron radiation, requires that the ratio $u_{ph}/u_B$ be higher than unity \citep{rybicki} where $u_B$ is the magnetic energy 
density and $u_{ph}$ is the radiation energy density due to the thermal emission at the dissipation site. The ratio can be rewritten as 
\begin{equation}
   \frac{u_{ph}}{u_B} = \frac{E_{ph}}{E_B} \equiv \frac{E_{BB}}{E_B} > 1   \label{eqn19}
\end{equation}
where $E_{B}$ is the total magnetic energy and $E_{BB}$ is the total energy of the BB component of 
the observed spectrum for the entire burst duration which is estimated to be around $4 
\times 10^{53}$ erg. Substituting this value into 
equation \ref{eqn19} suggests that $E_B$ is less than $4 \times 10^{53}$ erg, indicating that the total magnetic energy comprises less 
than $1\%$ of the total burst energy. Additionally, it suggests that the magnetic field intensity at the dissipation site, denoted by $B$, is less than approximately $10^{6}$ Gauss.
\\

Furthermore, the accelerated electrons in the shock region of the outflow cool on time-scales in the co-moving frame given by \( t'_{cool} \simeq 3 m_e c/4 \gamma_e \sigma_T u_{ph}\), where 
$\sigma_T$ is the Thomson cross-section. The thermal energy density at the dissipation radius, $R_d \sim 3.5 \times 10^{13}\, \rm cm$ 
is estimated to be  $u_{ph} = 2 \times 10^{12}$ erg/$cm^3$
Thus, the Compton cooling time-scale for the electrons is found to be around $ 10^{-8}$ s
which is much lesser than the dynamical time-scale, $t'_{dyn}$ = $R_d$/$\Gamma$c = 2.7 s, indicating that the Compton cooling is much efficient during the prompt emission of the burst. \\

\subsection{Insignificance of higher order ICS}
\label{Highorder_scattering}
 
Based on the calculations conducted in the preceding sections, we find that the kinetic energy dissipation of the jet occurs  within the optically 
thin region spanning from $10^{13}$ to $10^{14}$ cm (Figure \ref{fig:5}a). Although the outflow progresses to a region where the probability of photon-electron scattering diminishes, there 
can still be significant opacity for photon-photon interactions. 

The compactness parameter, $\ell$, which characterises the opacity for photon-photon annihilation is given by:

\begin{equation}
\ell = \frac{\sigma_T L_{\gamma ,> 500\hspace{0.2em}keV}}{8 \pi m_e R c^3 \Gamma^3}
\label{compactness}
\end{equation}

where $L_{\gamma , >500\hspace{0.2em}keV}$ represents the observed luminosity with the photons of energy exceeding $\sim 500$ keV ($\sim$ the 
rest-mass energy of an 
electron) in the co-moving frame, $\sigma_T$ is the Thomson cross-section, $m_e$ is the electron mass, $c$ is the speed of light, $R$ is the 
characteristic radius measured from the center of the GRB source, $\Gamma$ is the bulk Lorentz factor of the outflow. 

Considering the first-order scattered component of the time-integrated observed spectrum, we note that for a bulk Lorentz factor of $\Gamma = 434$, a threshold energy of 500 keV in the co-moving frame 
corresponds to $\sim 200$ MeV in the observer (lab) frame. For the time-integrated spectrum, we estimate the luminosity $L_{\gamma , >500\hspace{0.2em}keV}$ in the energy range 200 MeV to 5 GeV to be $1.86 \times 10^{52} \, erg\hspace{0.2em}\text{s}^{-1}$. This corresponds to a total number of photons exceeding 500 keV in the co-moving frame (or greater than 200 MeV in observer frame) which is given by:

\begin{equation}
N_{\gamma, >500 \text{ keV}} = \dot{N}_{\gamma, >500 \text{ keV}} \times \Delta t_{obs}
\end{equation}

where  $\dot{N}_{\gamma, >500 \text{ keV}} \sim \frac{L_{\gamma ,> 500\hspace{0.2em}keV}}{\Gamma m_e c^2 }$, $ \Delta t_{obs} = 8.6$ s is the duration of the interval and the estimate is $N_{\gamma, >500 \text{ keV}} = 4.5 \times 10^{56}$.
Using the estimate of $L_{\gamma , >500\hspace{0.2em}keV}$  and $\Gamma = 434$ in Equation \ref{compactness}, we find that the compactness parameter, $\ell$, remains greater than 1 up to a radius of $\ge 10^{14}$ cm. Since the dissipation site ($R_d$) lies within the region of $ < 10^{14}$ cm, where $\ell > 1$, photon-photon pair production (pp) is expected to occur. 

It is important to note that this estimated luminosity and the number of photons above 500 keV (in comoving frame)
are based on the observed spectrum. 
If significant pair production had 
occurred, the actual number of photons exceeding 500 keV in the co-moving frame would be higher, accounting for 
both the observed photons ($> 500$ keV in the co-moving frame) and the pairs created. 

The presence of a high-energy spectral tail extending into the GeV range suggests that, following pair 
production, the increased optical depth at the dissipation site must remain at or below 1. If the optical depth were to increase 
beyond 1 post-pair production, we would expect a turnover in the power-law segment due to thermalisation effects \citep{Ahlgren_etal_2019}. Since no such turnover is 
observed, we infer that the optical depth at the dissipation site remains at or below 1 after pair production. In other words, no significant modification in spectral shape is expected after pair production.

The optical depth for gamma-ray photon-electron scattering at distance $R$ is given by:

\begin{equation}
\tau_{\gamma e} = \frac{\dot{N_e} \sigma_T}{8 \pi \Gamma^2 c R}
\end{equation}
where $N_e = \dot{N_e} \times \Delta t_{obs}$ is the number of electrons and $\dot{N_e} = \frac{L}{\Gamma m_p c^2}$. Thus, for a given time interval at the dissipation site, the optical depth $\tau_{\gamma,e}$ is proportional to $N_e$. 

At $R_d$, the pre-pair-production (pre-pp) optical depth is found to be approximately $0.4$ (Figure \ref{fig:6}b). Assuming that post-pair production (post-pp), the optical depth increases to unity, we estimate the number of generated pairs, $N_{e,pp}$, as follows:

\begin{align}
N_e^{\text{post-pp}} &= \left( \frac{\tau^{\text{post-pp}} = 1}{\tau^{\text{pre-pp}}} \right) \times N_e^{\text{pre-pp}} \\
N_{e,pp} &= \frac{N_e^{\text{post-pp}} - N_e^{\text{pre-pp}}}{2}.
\end{align}

The number of generated pairs is found to be approximately $3.8 \times 10^{55}$. Since, only the photons above 500 keV interact to give rise to pair production, the total number of photons 
above 500 keV in the co-moving frame is $4.5 \times 10^{56}$ (observed photons of the time-integrated spectrum) $+ 2 \times 3.8 \times 10^{55} = 5.26 \times 10^{56}$. Thus, nearly $14\%$ of photons greater than 500 keV in the co-moving frame is converted into pairs.    

Since the total energy in electrons remains constant, 
the increase in 
leptonic particles post-pair production reduces the average Lorentz factor of electrons, $ \langle \gamma_e  \rangle$, from around $\sim 1400$ (as discussed in Section \ref{average_electron}) to around $\sim 500$ using equation \ref{eqn_avg_egamma}.

Within the overall electron distribution, the thermal component predominantly consists of electrons with $\gamma_e < 3$. Thus, the average $\gamma_e$ is primarily influenced by the power-law segment of the electron distribution. A reduction in the average 
$ \langle \gamma_e  \rangle$ can result from a steepening of the electron power-law index and a shift in the minimum electron 
Lorentz factor, $\gamma_{\min}$, towards the thermal part of the distribution. In both cases, lower values of $\gamma_e$ contribute more significantly to the average, leading to a decrease in the $ \langle \gamma_e  \rangle$.

A decrease in $\gamma_{min}$ can lead to a reduction in the thermal Lorentz factor $\gamma_{e,th}$ of the electron distribution. Since the electrons cool efficiently via inverse Compton scattering (see Section 
\ref{average_electron}) within the dynamical timescale—i.e., the time it takes for the outflow to exit the 
dissipation site—this results in the second-order Compton-scattered peak at lower energies and become less pronounced in energy flux than expected. Consequently, the 
contribution of the second-order scattered peak to the total observed flux with respect to the first-order scattering is expected to be less prominent.

Thus, in the proposed scenario, where pair production occurs at the dissipation site where $\ell > 1$, we anticipate that higher-order Compton scatterings will be less significant. 
This conclusion aligns with the observed spectral features, supporting the hypothesis that 
pair production plays a crucial role in shaping the optically thin inverse Compton scattered emission spectra of GRBs.

\section{Summary \& Conclusion}   \label{conclusion}
In this study, we investigate in detail the radiation mechanism of the prompt emission of GRB131014A. The detailed time resolved spectral analysis of the GRB revealed that the spectrum is more complex than the typical Band function. The spectrum has an unconventional shape with three spectral peaks and an extended high energy power law tail, best fitted by 2Bands + blackbody model with high statistical significance. 
The overall shape of the spectrum is characterized by five parameters namely lower energy power law index $\alpha$, higher energy power law index $\beta$ and the three spectral peaks $E_{peak1}$, $E_{peak2}$ and $E_{peak3}$. Such a complex spectral shape is reported for the first time. 

The persistently hard $\alpha$ values, surpassing $-0.5$ and occasionally reaching up to $+1$, defy an explanation through 
synchrotron radiation. Moreover, the presence of an extended power law extending to a few GeV, rather than exhibiting a cutoff beyond the highest spectral peak, contradicts localised sub-photospheric 
dissipation models. Therefore, the overall spectral shape, featuring an extended power law, is more likely consistent with first-order inverse Compton scattering. In this scenario, thermal photons advected 
from the photosphere undergo upscattering by electrons accelerated in the shock region within the optically thin outflow. The resulting spectrum reflects the shape of the post-shock electron distribution, which comprises both a thermal component and a power-law component.

Within this physical framework, we characterize the properties of inverse Compton scattering, including the amplification factor and 
Compton $y$ parameter, and determine electron distribution parameters, such as the electron power-law index, which is around $1.5$. Despite multiple emission episodes occurring throughout the burst duration, the Lorentz factor of the outflow steadily decreases over 
time. Concurrently, the nozzle radius of the jet initially expands from nearly $10^6$ cm to approximately $10^9$ cm before becoming steady. This temporal behavior of the outflow aligns with the 
observations of single-pulsed emission GRBs. The shock region is estimated to lie just above the photosphere at an optical depth of $0.4$. Furthermore, our analysis indicates that approximately $76\%$ 
of the total burst energy goes into electrons, while less than $1\%$ goes to magnetic fields. Moreover, given that the dissipation site is well within the photon compactness region, we suggest that a portion of the upscattered photons in the first-order scattering, exceeding 500 keV in the co-moving frame, undergo pair production, ensuring that the resulting optical depth 
remains unity. This pair production is anticipated to lead to a decrease in the average electron Lorentz factor across both the thermal and power-law components of the electron 
distribution. Consequently, higher-order scatterings are suppressed, having a minimal impact on the observed first-order scattered inverse Compton spectrum.

In conclusion, this study offers a thorough identification and detailed modeling of optically thin inverse Compton scattering in the prompt emission of GRB 131014A. A more robust validation of this interpretation, through direct physical model fitting of the optically thin inverse Compton scattering framework to the GRB data, will be pursued and presented in a dedicated future work. 

\appendix

\section{Fit parameters of the best fit model \& Other details}
\label{appendix1}
In Table \ref{table_fit}, the fit values for the eight free parameters of the best-fit model of `2 Bands' up to 1.49 s, and the ten free parameters of the best-fit model of `2 Bands + Blackbody' above 1.49 s, are listed. Additionally, the $\Delta$ AIC ($\rm AIC_{\rm model} - \rm AIC_{\rm best \, model}$) values for the various models relative to the best-fit model are also provided for all the time resolved intervals.  If the $\rm AIC_{\rm model}$ for a given model results in $\Delta \rm AIC < 4$, the model is regarded as equally plausible compared to the best-fit model. In contrast, models with $\Delta \rm AIC \gg 4$ are deemed implausible, affirming that the selected best-fit model is indeed the most appropriate \citep{AIC_selection2011}. Furthermore, we note that AIC inherently penalises more complex models, yet in this study we find 2Bands + BB (or 2 Bands only in  0 s to 1.49 s)\footnote{In the time bins 1.49–2.22 s and 2.7-3.04 s, we find that both the 2Bands and 2Bands + BB models provide equally good fits with comparable statistical significance and the comparison yields $\Delta AIC = 0$, indicating no statistical preference between the models. 
However, we successfully fit the 2Bands + BB model and obtain well-constrained parameters for the BB component. The equally good fit suggests that this model is just as plausible, with the derived BB 
parameters being consistent with the evolutionary trend observed in later time bins. Furthermore, after accounting for the additional parameters in 
2Bands + BB, the AIC remains equal to that of the 2Bands-only model. Therefore, we adopt 2Bands + BB as the best-fit model for this interval, as it offers greater physical consistency.} consistently yields the lowest AIC among all the models tested indicating they are well-consistent with the data. 

\begin{figure}[!ht]
    \centering
    \includegraphics[width=1.0\textwidth]{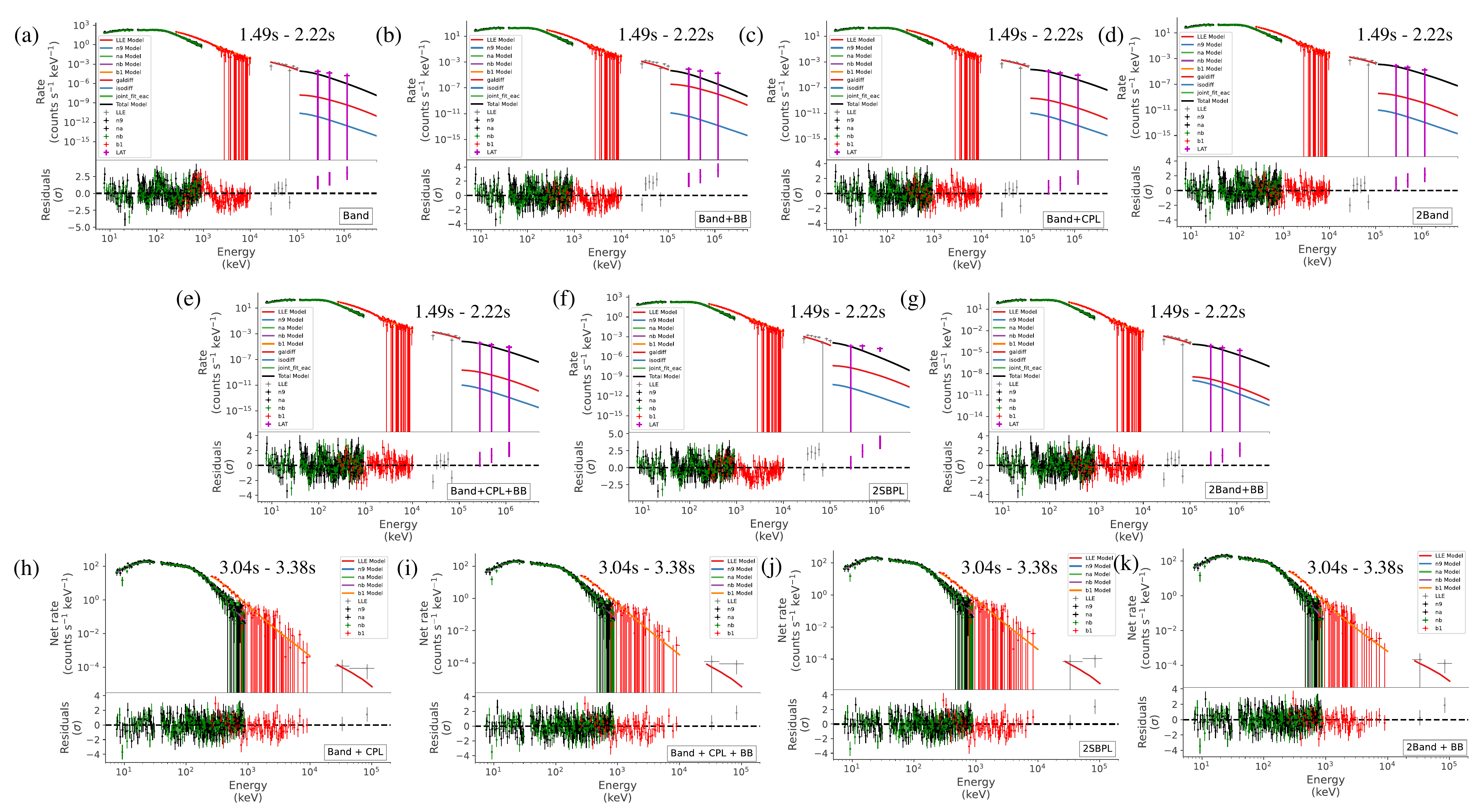}
    \caption{The counts plot along with the residuals  obtained for the different spectral models: (a) Band, (b) Band + BB, (c) Band + CPL, (d) 2Bands, (e) Band + CPL + BB, (f) 2SBPL and (g) 2 Bands + BB are shown for the time interval [1.49 s - 2.22 s] where both LLE and LAT data are present. The counts plot along with the residuals obtained for the spectral models: (h) Band + CPL, (i) Band + CPL + BB, (j) 2SBPL model and the best fit model, (k) 2 Bands + BB in the time interval [3.04 s - 3.38 s] are shown.} 
    \label{Residual_2}
\end{figure}

A significant LLE and LAT emission begins around 1.49 s.  In Figure \ref{Residual_2} (a-g), the counts and residual plots obtained for the different models in the time interval [1.49s - 2.22s] reported in Table \ref{table_fit} were LLE and LAT data are present are shown.
In Figure \ref{Residual_2} (h-k) the residuals of various spectral models tested with the GRB data in its bright bin [3.04s - 3.38s] which includes Band + Cutoff Powerlaw, 2Band, Band + Cutoff Powerlaw + Blackbody, and 2SBPL, in comparison with the best-fit model, 2Bands + Blackbody, are presented.
\\
In addition, the functional form of the various spectral  models used in this study are given below:

(1) The Band function is defined as
\[
N(E) =
\begin{cases} 
A \left(\frac{E}{100}\right)^\alpha \exp\left(-\frac{E}{E_0}\right), & \text{if } E \leq (\alpha - \beta) E_0, \\
A \left[\frac{(\alpha - \beta) E_0}{100}\right]^{\alpha - \beta} \exp(\beta - \alpha) \left(\frac{E}{100}\right)^\beta, & \text{if } E > (\alpha - \beta) E_0.
\end{cases}
\]

where,
\begin{itemize}
    \item $A$ is the normalisation constant,
    \item $\alpha$ and $\beta$ are the low-energy and high-energy photon indices,
    \item $E_0$ is the break energy.
    \item The spectral peak $E_{peak}$ = $E_0$($\alpha$ + 2).
\end{itemize} 

\vline\

\vline\

(2) The Smoothly Broken Power Law (SBPL) function is defined as
\[
N(E) = A \left(\frac{E}{E_0}\right)^{\alpha} \left[1 + \left(\frac{E}{E_b}\right)^{\Delta}\right]^{\frac{\beta - \alpha}{\Delta}}
\]
where,
\begin{itemize}
    \item $A$ is the normalisation constant,
    \item $E_0$ is the pivot energy (often set to 100 keV),
    \item $E_b$ is the break energy,
    \item $\alpha$ is the low-energy photon index,
    \item $\beta$ is the high-energy photon index,
    \item $\Delta$ controls the smoothness of the break.
\end{itemize}

\vline\

\vline\

(3) The Cutoff Power Law (CPL) function is defined as:
\[
N(E) = A \left(\frac{E}{E_0}\right)^\alpha \exp\left(-\frac{E}{E_c}\right)
\]
where:
\begin{itemize}
    \item $A$ is the normalisation constant,
    \item $E_0$ is the pivot energy (often set to 100 keV),
    \item $\alpha$ is the low-energy photon index,
    \item $E_c$ is the cutoff energy.
\end{itemize}

\vline\

\vline\

(4) The Blackbody (BB) function for GRB spectrum is defined as:
\[
N(E) = K \frac{E^2}{\exp\left(\frac{E}{kT}\right) - 1}
\]

where:
\begin{itemize}
    \item $E$ is the photon energy,
    \item $K$ is the normalisation constant,
    \item $k$ is Boltzmann's constant,
    \item $T$ is the temperature of the blackbody in Kelvin.
\end{itemize}
\vline\

\vline\

(5) The 2SBPL function for GRB spectrum is defined as:
\\

\[
N_E^{\text{2SBPL}} = A E_{\text{break}}^{\alpha_1} 
\left[ 
    \left[ 
        \left( \frac{E}{E_{\text{break}}} \right)^{-\alpha_1 n_1} 
        + \left( \frac{E}{E_{\text{break}}} \right)^{-\alpha_2 n_1} 
    \right]^{\frac{n_2}{n_1}} 
    +  \left( \frac{E}{E_j} \right)^{-\beta n_2} 
    \cdot 
    \left[ 
        \left( \frac{E_j}{E_{\text{break}}} \right)^{-\alpha_1 n_1} 
        + \left( \frac{E_j}{E_{\text{break}}} \right)^{-\alpha_2 n_1} 
    \right]^{\frac{n_2}{n_1}} 
\right]^{- \frac{1}{n_2}}
\]
\\

And,
\[
E_j = E_{\text{peak}} \cdot 
\left( \frac{-\alpha_2 + 2}{\beta + 2} \right)^{\frac{1}{(\beta - \alpha_2) n_2}}
\]

where:
\begin{itemize}
    \item $E$ is the photon energy,
    \item $A$ is the normalisation constant,
    \item $E_{break}$ is the break energy,
    \item $E_{peak}$ is the peak energy,
    \item $\alpha_1$ is the photon index below $E_{break}$,
    \item $\alpha_2$ is the photon index between $E_{break}$ and $E_{peak}$,
    \item $\beta$ is the high energy photon index above $E_{peak}$,
    \item $n_1$ is the smoothness parameter for $E_{break}$,
    \item $n_2$ is the smoothness parameter for $E_{peak}$,
\end{itemize}
\vline\

\begin{sidewaystable*}
\centering
\setlength{\tabcolsep}{1pt}
\renewcommand{\arraystretch}{1.0}
\fontsize{05}{13}\selectfont 
    \begin{tabular}{|c|c|c|c|c|c|c|c|c|c|c|c|c|c|c|c|c|}
    \hline
    \multicolumn{17}{|c|}{\normalsize \textbf{Time-resolved analysis results}}
    \\ \hline
    \multicolumn{1}{|c|}{\normalsize Time intervals} & \multicolumn{2}{c|}{\normalsize Blackbody} & \multicolumn{4}{c|}{\normalsize Band$_1$} & \multicolumn{4}{c|}{\normalsize Band$_2$} & \multicolumn{6}{|c|}{\normalsize $\Delta$AIC = AIC$_{model}$ - AIC$_{best\, model}$} \\ \cline{2-17} 
    \scriptsize (second) & \normalsize K & \normalsize kT & \normalsize K$_1$ & \normalsize $\alpha_1$ & \normalsize $\beta_1$ & \normalsize Epeak$_1$ & \normalsize K$_2$ & \normalsize $\alpha_2$ & \normalsize $\beta_2$ & \normalsize Epeak$_2$ & \scriptsize $Band$ & \scriptsize $Band+$ & \scriptsize $Band+ $ & \scriptsize $2 Band$ & \scriptsize $Band+$ & \scriptsize $2SBPL$ \\ 
     & (cm$^{-2}$keV$^{-3}$s$^{-1}$) & \scriptsize (keV) & (cm$^{-2}$keV$^{-1}$s$^{-1}$) &  &   & \scriptsize (keV) & (cm$^{-2}$keV$^{-1}$s$^{-1}$) &   &   & \scriptsize (keV) &   & \scriptsize BB  & \scriptsize CPL  &   &  \scriptsize CPL+ BB &   \\ \hline
    \scriptsize 0.00 - 0.95 & \scriptsize None & \scriptsize None & 0.35$^{+0.15}_{-0.17}$ & 0.80$^{+0.62}_{-0.67}$ & -2.55$^{+0.16}_{-0.15}$ & 164$^{+23}_{-21}$ & 0.09$^{+0.02}_{-0.02}$ & -0.44$^{+0.13}_{-0.13}$ & -4.99$^{+0.04}_{-0.01}$ & 440$^{+51}_{-61}$ & \scriptsize4 & \scriptsize1 & \scriptsize0 & \scriptsize0 & \scriptsize None & \scriptsize6 \\ \hline
    \scriptsize0.95 - 1.09 & \scriptsize None & \scriptsize None & 1.04$^{+0.54}_{-0.51}$ & 0.80$^{+0.64}_{-0.62}$ & -3.09$^{+0.91}_{-1.15}$ & 115$^{+16}_{-17}$ & 0.26$^{+0.04}_{-0.04}$ & -0.24$^{+0.20}_{-0.20}$ & -4.99$^{+0.19}_{-0.03}$ & 410$^{+30}_{-29}$ & \scriptsize4 & \scriptsize3 & \scriptsize2 & \scriptsize0 & \scriptsize None & \scriptsize1 \\ \hline
    \scriptsize1.09 - 1.23 & \scriptsize None & \scriptsize None & 1.11$^{+0.89}_{-0.65}$ & 1.36$^{+0.91}_{-0.73}$ & -3.17$^{+0.89}_{-0.99}$ & 150$^{+23}_{-29}$ & 0.43$^{+0.07}_{-0.07}$ & -0.35$^{+0.10}_{-0.08}$ & -4.96$^{+0.66}_{-0.15}$ & 399$^{+32}_{-31}$ & \scriptsize4 & \scriptsize4 & \scriptsize2 & \scriptsize0 & \scriptsize None & \scriptsize2 \\ \hline
    \scriptsize1.23 - 1.34 & \scriptsize None & \scriptsize None & 0.79$^{+0.70}_{-0.65}$ & 3.01$^{+0.001}_{-0.002}$ & -3.24$^{+1.00}_{-1.14}$ & 194$^{+30}_{-30}$ & 0.64$^{+0.09}_{-0.08}$ & -0.20$^{+0.09}_{-0.09}$ & -3.45$^{+0.48}_{-0.46}$ & 449$^{+36}_{-37}$ & \scriptsize3 & \scriptsize2 & \scriptsize4 & \scriptsize0 & \scriptsize None & \scriptsize3 \\ \hline
    \scriptsize1.34 - 1.44 & \scriptsize None & \scriptsize None & 3.44$^{+1.02}_{-1.16}$ & 3.00$^{+0.005}_{-0.003}$ & -2.55$^{+0.35}_{-0.25}$ & 161$^{+12}_{-11}$ & 0.31$^{+0.05}_{-0.05}$ & -0.60$^{+0.09}_{-0.09}$ & -2.87$^{+0.28}_{-0.33}$ & 525$^{+64}_{-69}$ & \scriptsize1 & \scriptsize3 & \scriptsize1 & \scriptsize0 & \scriptsize None & \scriptsize5  \\ \hline
    \scriptsize 1.44 - 1.49 & \scriptsize None & \scriptsize None & 2.20$^{+1.01}_{-1.12}$ & 2.99$^{+0.001}_{-0.01}$ & -2.85$^{+0.59}_{-0.74}$ & 188$^{+24}_{-25}$ & 0.56$^{+0.06}_{-0.05}$ & -0.24$^{+0.08}_{-0.06}$ & -4.07$^{+1.25}_{-0.08}$ & 687$^{+47}_{-61}$ & \scriptsize1 & \scriptsize1 & \scriptsize2 & \scriptsize0 & \scriptsize None & \scriptsize3 \\ \hline
    \scriptsize 1.49 - 2.22 & 0.0023$^{+0.002}_{-0.001}$ & 13.47$^{+2.34}_{-2.45}$ & 2.15$^{+0.61}_{-0.54}$ & 1.36$^{+0.56}_{-0.51}$ & -2.44$^{+0.04}_{-0.04}$ & 132$^{+19}_{-22}$ & 0.80$^{+0.08}_{-0.08}$ & -0.26$^{+0.06}_{-0.05}$ & -3.75$^{+0.44}_{-0.39}$ & 606$^{+23}_{-25}$ & \scriptsize137 & \scriptsize20 & \scriptsize3 & \scriptsize0 & \scriptsize4 & \scriptsize15 \\ \hline
    \scriptsize2.22 - 2.47 & 0.003$^{+0.002}_{-0.0008}$ & 15.26$^{+0.61}_{-4.67}$ & 8.33$^{+6.35}_{-3.52}$ & 2.99$^{+0.07}_{-0.51}$ & -2.29$^{+0.07}_{-0.10}$ & 137$^{+7}_{-26}$ & 0.52$^{+0.11}_{-0.10}$ & -0.46$^{+0.12}_{-0.06}$ & -4.28$^{+1.77}_{-0.29}$ & 277$^{+15}_{-20}$ & \scriptsize6 & \scriptsize2 & \scriptsize2 & \scriptsize3 & \scriptsize None & \scriptsize9 \\ \hline
    \scriptsize2.47 - 2.70 & 0.005$^{+0.004}_{-0.001}$ & 15.30$^{+1.38}_{-7.04}$ & 7.70$^{+2.22}_{-6.62}$ & 2.55$^{+0.07}_{-1.82}$ & -2.74$^{+0.33}_{-0.30}$ & 161$^{+14}_{-22}$ & 0.47$^{+0.08}_{-0.13}$ & -0.40$^{+0.19}_{-0.13}$ & -3.19$^{+0.44}_{-0.52}$ & 448$^{+50}_{-43}$ & \scriptsize20 & \scriptsize3 & \scriptsize4 & \scriptsize4 & \scriptsize1 & \scriptsize3 \\ \hline
    \scriptsize2.70 - 3.04 & 0.004$^{+0.001}_{-0.001}$ & 13.20$^{+1.11}_{-1.12}$ & 6.82$^{+1.11}_{-1.25}$ & 1.90$^{+0.25}_{-0.25}$ & -2.44$^{+0.04}_{-0.03}$ & 150$^{+1}_{-1}$ & 0.59$^{+0.05}_{-0.05}$ & -0.58$^{+0.05}_{-0.05}$ & -4.99$^{+0.32}_{-0.04}$ & 286$^{+1}_{-1}$ & \scriptsize25 & \scriptsize9 & \scriptsize2 & \scriptsize0 & \scriptsize1 & \scriptsize6 \\ \hline
    \scriptsize3.04 - 3.38 & 0.02$^{+0.002}_{-0.001}$ & 11.85$^{+0.38}_{-0.61}$ & 25.62$^{+4.18}_{-3.48}$ & 3.00$^{+0.002}_{-0.002}$ & -2.67$^{+0.09}_{-0.08}$ & 134$^{+3}_{-8}$ & 0.19$^{+0.16}_{-0.04}$ & 1.97$^{+0.38}_{-2.09}$ & -5.00$^{+0.002}_{-0.001}$ & 356$^{+15}_{-28}$ & \scriptsize49 & \scriptsize28 & \scriptsize26 & \scriptsize23 & \scriptsize16 & \scriptsize31  \\ \hline
    \scriptsize3.38 - 3.69 & 0.02$^{+0.003}_{-0.002}$ & 11.27$^{+0.43}_{-0.68}$ & 31.58$^{+6.27}_{-3.44}$ & 3.00$^{+0.001}_{-0.001}$ & -2.50$^{+0.06}_{-0.04}$ & 132$^{+2}_{-9}$ & 0.23$^{+0.16}_{-0.01}$ & 2.78$^{+0.13}_{-1.97}$ & -4.10$^{+0.78}_{-0.47}$ & 345$^{+14}_{-20}$ & \scriptsize42 & \scriptsize23 & \scriptsize16 & \scriptsize13 & \scriptsize5 & \scriptsize21 \\ \hline
    \scriptsize3.69 - 4.65 & 0.01$^{+0.004}_{-0.003}$ & 11.73$^{+1.17}_{-1.32}$ & 14.39$^{+5.06}_{-4.74}$ & 2.99$^{+0.005}_{-0.04}$ & -2.43$^{+0.06}_{-0.07}$ & 127$^{+9}_{-10}$ & 0.29$^{+0.06}_{-0.07}$ & -0.74$^{+0.11}_{-0.10}$ & -5.00$^{+0.003}_{-0.001}$ & 260$^{+27}_{-23}$ & \scriptsize 16 & \scriptsize12 & \scriptsize6 & \scriptsize6 & \scriptsize9 & \scriptsize5  \\ \hline
    \scriptsize4.65 - 8.61 & 0.003$^{+0.001}_{-0.0002}$ & 7.74$^{+0.10}_{-0.10}$ & 0.04$^{+0.005}_{-0.005}$ & -1.06$^{+0.06}_{-0.12}$ & -4.99$^{+0.54}_{-0.06}$ & 82$^{+0.5}_{-0.5}$ & 0.01$^{+0.001}_{-0.007}$ & -1.11$^{+0.28}_{-0.26}$ & -2.29$^{+0.21}_{-0.14}$ & 225$^{+106}_{-10}$ &  \scriptsize 4 &  \scriptsize 4 &  \scriptsize 3 &  \scriptsize 2 &  \scriptsize 1 &  \scriptsize 5  \\ \hline
    \multicolumn{17}{|c|}{ \normalsize \textbf{Time-integrated analysis results}}
    \\ \hline
    \scriptsize0.00 - 8.61 & 0.002$^{+0.0005}_{-0.0005}$ & 12.39$^{+0.62}_{-0.62}$ & 3.41$^{+0.42}_{-0.41}$ & 2.99$^{+0.001}_{-0.001}$ & -2.39$^{+0.03}_{-0.03}$ & 150.31$^{+4.28}_{-4.23}$ & 0.24$^{+0.01}_{-0.01}$ & -0.55$^{+0.03}_{-0.03}$ & -3.33$^{+0.23}_{-0.22}$ & 451.34$^{+11.63}_{-12.29}$ & \scriptsize 229 &  \scriptsize 79 & \scriptsize 79 &  \scriptsize 59 &  \scriptsize 30 &  \scriptsize 97  \\ \hline
    \end{tabular}
\caption{The fit parameters and its errors of the spectral components of the best fit model along with the AIC difference of the different models with respect to the best fit model are listed. In the time-resolved spectral analysis, in the initial 6 time intervals, 2Bands model is considered to be the best fit model while in the remaining 8 time intervals, 2Band + BB is the best fit model. 'None' indicates that the model could not be fitted to the data. In time-integrated spectral analysis, the best fit model is found to be 2Band + BB.}
\label{table_fit}
\end{sidewaystable*}

\section{Comparison with Simulated Inverse Compton spectrum}
\label{appendix2}
\begin{figure}[!ht]
    \centering
    \includegraphics[scale =0.22]{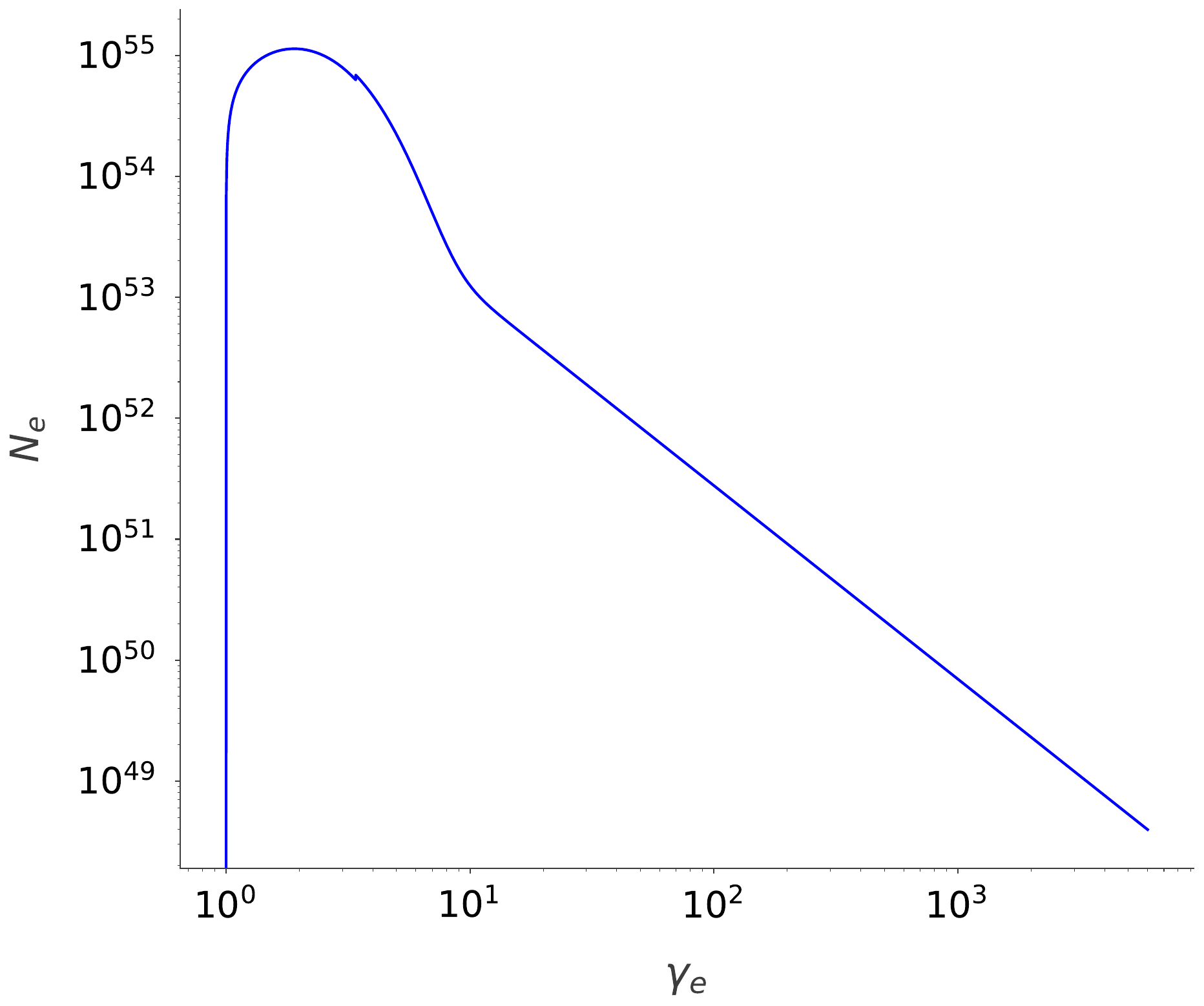}
    \caption{Post-shock electron distribution with $\delta$=1.65, $\theta$= 0.8, $\gamma_{min}$=3.24, $n_0$=3.00$\times$$10^{55}$ is shown.}    
    \label{simulated_elec_dist}
\end{figure}

\begin{figure*}[!ht]
    \centering
    \subfigure[Observed spectrum]{
        \includegraphics[width=0.395\textwidth]{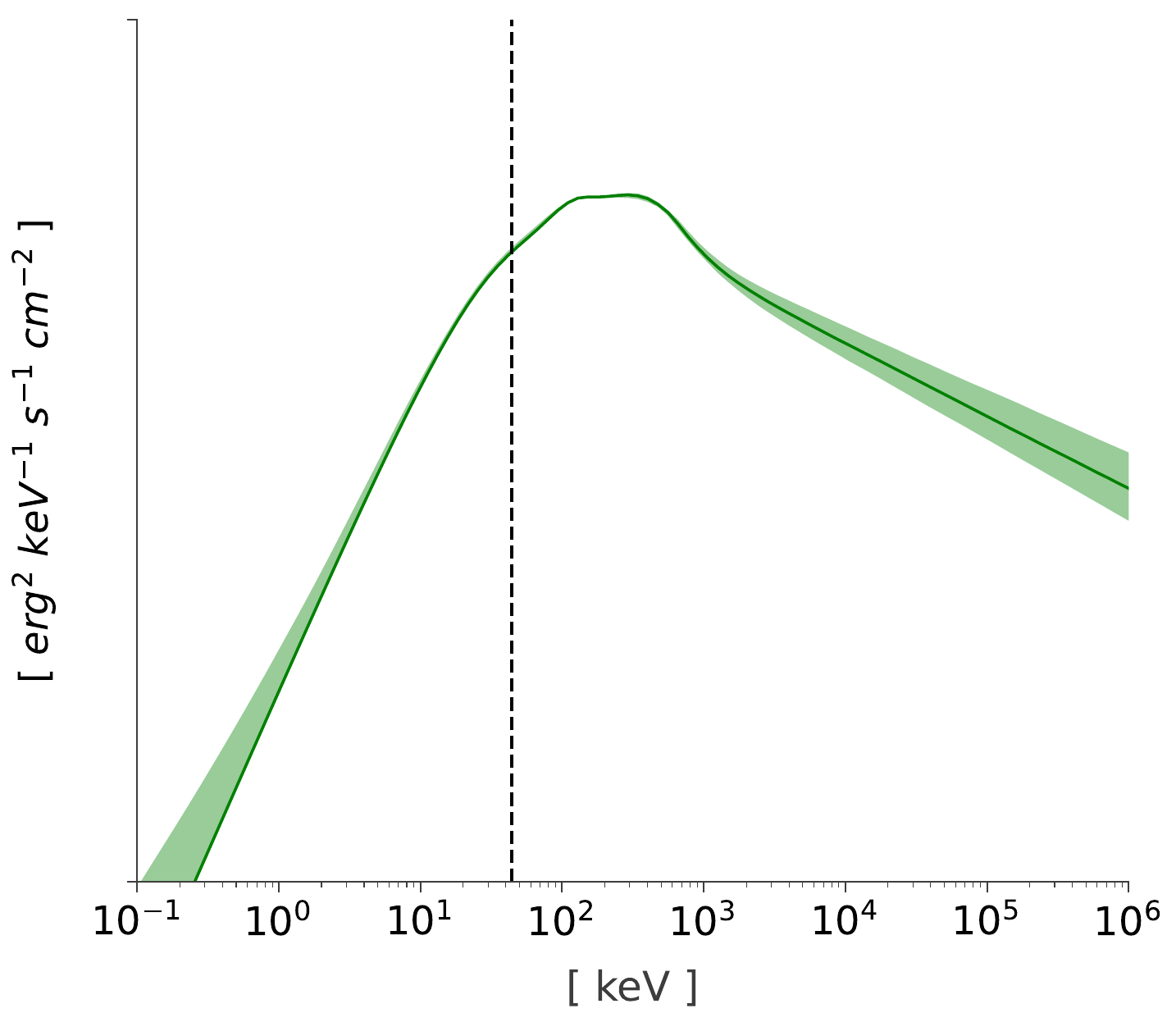}
    }
    \hspace{0.15in}
    \subfigure[Simulated spectrum]{
        \includegraphics[width=0.395\textwidth]{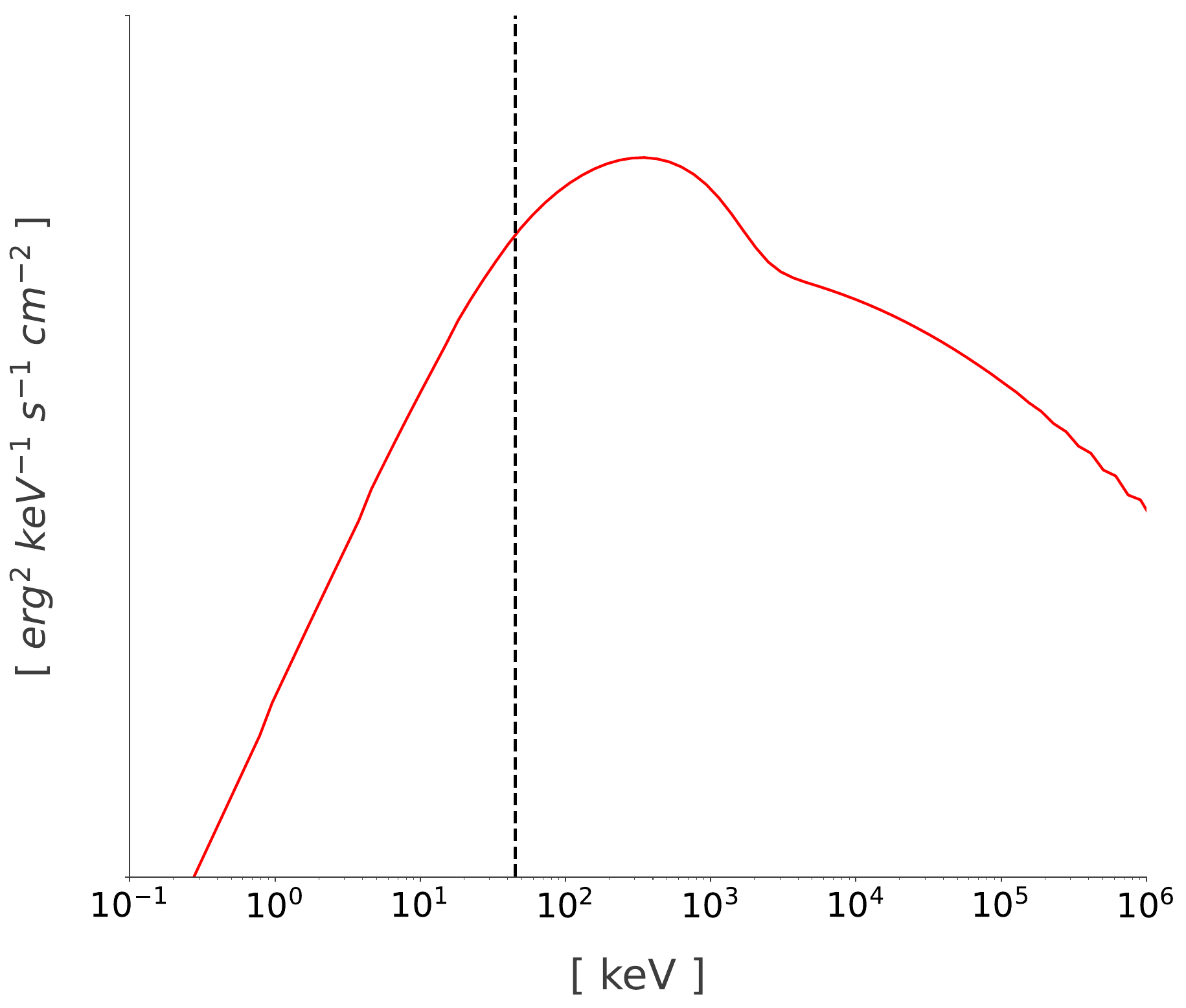}
    }
    \caption{ Comparison between the observed and simulated spectra. (a) Observed spectrum best fitted by 2Bands + Blackbody model is shown in green solid line. (b) Simulated spectrum of first order inverse Compton scattering is shown. The black dashed vertical lines represent blackbody temperature at 12 keV. }
    \label{fig:simulated spectrum}
\end{figure*}

Using the electron distribution characteristics obtained in section \ref{ICS3} for one of the brighter time intervals, i.e 
[3.04s - 3.38s] (Figure \ref{simulated_elec_dist}) and the blackbody function of temperature 12 keV (observer frame) as the seed photon 
distribution, a first order inverse Compton scattering spectrum is simulated using the Naima 
package \citep{naima}. The obtained simulated observed spectrum is shown in Figure \ref{fig:simulated spectrum} is found to closely resemble the observed 
spectrum of the same interval shown on the left in the Figure \ref{fig:simulated spectrum}. 
This preliminary analysis provides encouraging results supporting the physical scenario that the observed spectrum is a result of first-order optically thin inverse Compton scattering of photospheric emission. In a subsequent future work, we aim to directly model the GRB data with the ICS model using the Naima package.    

\begin{acknowledgments}
We thank the referee for their insightful suggestions. S.I. is supported by DST INSPIRE Faculty Scheme (IFA19-PH245) and SERB SRG Grant (SRG/2022/000211). This
research has made use of {\it Fermi} data obtained through High Energy Astrophysics Science Archive Research Center Online Service, provided by the NASA/Goddard Space Flight Center. This work utilized various software such as 3ML, NAIMA, PYTHON (\citealt{Python_1}), ASTROPY (\citealt{Astropy_2013,Astropy_2018,Astropy2022}), NUMPY (\citealt{Numpy2011,Numpy_1}), SCIPY (\citealt{Scipy2001,Scipy_1}), MATPLOTLIB (\citealt{Matplotlib_1}), FTOOLS (\citealt{ftools_1}) etc. We acknowledge the support of High Performance Computing Centre (CHPC) of IISER TVM for providing the computational resources. 
\end{acknowledgments}

\newpage
\bibliography{Ref1}

\end{document}